\documentclass[runningheads,envcountsame]{llncs}
\usepackage[T1]{fontenc}
\usepackage{graphicx}
\usepackage{tikz}
\usetikzlibrary{calc}

\usepackage{amsmath}
\usepackage{amssymb}

\usepackage{verbatim}
\usepackage[linesnumbered]{algorithm2e}

\spnewtheorem{claim*}[lemma]{Claim}{\itshape}{\normalfont}
\spnewtheorem*{cproof}{Proof of Claim}{\itshape}{\rmfamily}
\newenvironment{proof*}
{\begin{proof}
	}
	{\qed
\end{proof}}

\newcommand{\su}{^{\$}}
\newcommand{\rk}{\text{rk}}
\def \<#1>{{\langle {#1} \rangle}}
\newcommand\lcop[1]{\left\lfloor{#1}\right\rfloor}
\newcommand\paras[1]{\mathsf{ps}(#1)}
\newcommand\he[1]{\mathsf{ht}(#1)}

\newcommand\tree[1]{\mathcal{T}_{#1}}
\newcommand\emp{\emptyset}

\newcommand\ab{\allowbreak}

\newcommand{\QY}{\Theta}

\newcommand{\N}{\mathbb{N}}

\title{Deciding Linear Height and Linear Size-to-Height Increase of
  Macro Tree Transducers}
\titlerunning{Deciding Linear Increase Properties of Macro Tree Transducers}
\author{Paul Gallot\inst{1}\and Sebastian Maneth\inst{1} \and Keisuke Nakano\inst{2} \and Charles Peyrat\inst{3} }{}{}{}{}
\institute{Universit\"at Bremen, Germany \and Tohoku University, Sendai, Japan\and ENS Paris-Saclay, France }

\begin{document}

\maketitle

\begin{abstract}
We present a novel normal form for (total deterministic) macro tree transducers (mtts),
called ``depth proper normal form''. If an mtt is in this normal form,
then it is guaranteed that each parameter of each state 
%of the mtt 
appears 
at arbitrary depths in the output trees of that state. Intuitively, if some parameter
only appears at certain bounded depths in the output trees of a state, then
this parameter can be eliminated by in-lining the corresponding output paths at
each call site of that state. We use regular look-ahead in order to determine
which of the paths should be in-lined. As a consequence of changing the look-ahead,
a parameter that was previously appearing at unbounded depths,
may be appearing at bounded depths for some new look-ahead; for this reason, our
construction has to be iterated 
%in order 
to obtain an mtt in depth-normal form.
Using the normal form, we can decide whether the translation of an mtt
has linear height increase or has linear size-to-height increase.
\end{abstract}

\section{Introduction}

Tree transducers are fundamental devices in theoretical computer science.
%automata theory.
They generalize the finite state transductions from strings to (finite, ranked) trees and
were invented in the 1970s in the context of compiler theory and mathematical linguistics.
The most basic such transducers are the top-down tree
transducer~\cite{DBLP:journals/jcss/Thatcher70,DBLP:journals/mst/Rounds70}
and the bottom-up tree transducer~\cite{th70}, see also~\cite{DBLP:journals/mst/Engelfriet75}.
These transducers traverse their input tree once, but may process subtrees in several copies.
It is well known that these transducers have \emph{linear height increase} (``LHI''),
see e.g.~\cite{DBLP:series/eatcs/FulopV98}.

In this paper we deal with a more powerful kind of tree transducer:
the macro tree transducer~\cite{DBLP:journals/jcss/EngelfrietV85} (``mtt'').
Mtts can be seen as particularly simple functional
programs on trees restricted to primitive recursion via (input) tree pattern matching.
Alternatively, mtts can be seen as context-free tree grammars
(introduced in~\cite{DBLP:journals/mst/Rounds70} as ``context-free dendrogrammars'';
see also~\cite{fis68,DBLP:journals/jcss/EngelfrietS77,DBLP:journals/jcss/EngelfrietS78}
and~\cite[Section~15]{DBLP:reference/hfl/GecsegS97}),
the nonterminals of
which are controlled by a top-down tree storage (in the spirit of~\cite{DBLP:journals/corr/Engelfriet14}).
It is an open problem, if it is decidable for a given mtt whether or not
its translation can be realized by a top-down tree transducer
(with ``origin'' semantics, this is decidable~\cite{DBLP:journals/iandc/FiliotMRT18}).
As mentioned above, it is a necessary condition for the mtt to have
linear height increase (``LHI''). 
This raises the question, can we decide for a given
mtt, whether or not its translation has LHI?
Here we give an affirmative answer to this question.
---
It is also an open problem, if it is decidable for a given mtt whether or not
its translation can be realized by an attributed tree 
transducer~\cite{DBLP:journals/mst/Knuth68,DBLP:journals/actaC/Fulop81} (``att'').
It is well-known that atts have \emph{linear size-to-height increase} (``LSHI''), 
see, e.g.,~\cite{DBLP:series/eatcs/FulopV98}. 
This raises the question, can we decide for a given
mtt, whether or not its translation is of LSHI?
%Also to this question 
We give an affirmative answer.
Note that it was conjectured in~\cite{DBLP:journals/siamcomp/EngelfrietM03}
that the methods of that paper
could be adapted to give such an affirmative answer.

Let us now discuss our results in more detail.
To decide both the LHI and LSHI properties, we introduce a new normal form
called ``depth proper''. An mtt is depth proper if each parameter of every 
(reachable) state
appears at infinitely many different depths (for different input trees).
The idea of our construction is to eliminate 
parameters that only appear at bounded depths; we use regular
look-ahead to determine which bounded paths to output at a given moment.
Since in this way we may generate output ``earlier'' than the original transducer,
new ``helper states'' need to be introduced which continue the translation at
the correct input nodes. Both issues, the change of look-ahead and the introduction
of new states may cause the newly constructed transducer \emph{not} to be depth proper.
For this reason our construction has to be iterated.
% before a depth proper mtt
%is obtained.

To understand the idea of the construction, let us consider a simple example.
We consider input and output trees over a binary symbol $f$ and the
nullary symbol $a$ and an mtt with the following rules.
\[
\begin{array}{lcllcl}
q_0(f(x_1,x_2)) &\to& f(q_1(x_2), q_2(x_2, q_{\text{id}}(x_1))) &
  q_2(f(x_1,x_2),y_1) &\to& f(y_1, q_0(x_2))\\  
q_0(a)&\to& a &
  q_2(a,y_1) &\to& f(y_1, a)\\  
q_1(f(x_1,x_2)) &\to& q_{\text{id}}(x_1) &
  q_{\text{id}}(f(x_1,x_2)) &\to& f(q_{\text{id}}(x_1), q_{\text{id}}(x_2))\\  
q_1(a)&\to& a &
q_{\text{id}}(a) &\to& a
\end{array}
\]
The transducer realizes the following translation:
\[
f(t_1, f(t_2, \underbrace{f(t_3, f(t_4, \dots ))}_{t} )) \Rightarrow
f(t_2, f(t_1, f(t_4, f(t_3, \dots ))))
\]
Let us consider in detail how the rules of the mtt are applied 
to produce as output first the tree $t_2$ and then the tree $t_1$:
\[
\begin{array}{cl}
& q_0(f(t_1,f(t_2,t)))\\
\Rightarrow_{\text{first rule}} & f( q_1(f(t_2,t)), q_2(f(t_2,t), q_{\text{id}}(t_1)))\\
\Rightarrow_{\text{second rule}} & f( q_{\text{id}}(t_2), q_2(f(t_2,t), q_{\text{id}}(t_1)))\\
\Rightarrow_{\text{last two rules}}^* & f( t_2, q_2(f(t_2,t), q_{\text{id}}(t_1)))\\
\Rightarrow_{\text{third rule}} & f( t_2, f(q_{\text{id}}(t_1),q_0(t))\\
\Rightarrow_{\text{last two rules}}^* & f( t_2, f(t_1,q_0(t)))
\end{array}
\]  
%Let us consider the first (top-most) rule: the state $q_1$ produces
%the tree $t_2$. The state $q_2$ takes in its parameter argument the
%tree $t_1$ (produced by the call $q_{\text{id}}(x_1)$).
%At the \emph{next} node of the right-comb of the input tree
%the state $q_2$ outputs $f(t_1, t)$ where $t$ is the translation of the transducer 
%for the tree $f(t_3, \dots)$.
Parameters of an mtt are always instantiated by output trees.
In the example, the parameter $y_1$ of state $q_2$ is instantiated by the output
tree that is produced by the call $q_{\text{id}}(t_1)$.
Observe further, that state $q_2$ is 
\emph{not} depth proper: each tree that it outputs is of the form
$f(y_1, t)$ where $t$ does not contain the parameter $y_1$.
The idea of our construction is to replace each occurrence of state $q_2$ in
the right-hand side of any rule by this tree ``fragment'', where at the 
position of $t$ there will be the new ``helper state'' $[q_2,2]$. 
The path ``2'' indicates that this state should produce the tree at the second
child position of the output tree produced by $q_2$.
We obtain the following (the rules of $q_0,q_1$ and input $a$ are as before):
\[
\begin{array}{lcl}
q_0(f(x_1,x_2)) &\to& f(q_1(x_2), f(q_{\text{id}}(x_1), [q_2,2](x_2) )) \\
q_1(f(x_1,x_2)) &\to& q_{\text{id}}(x_1)\\
{[q_2,2]}(f(x_1,x_2)) &\to& q_0(x_2)\\
{[q_2,2]}(a) &\to& a\\
q_{\text{id}}(f(x_1,x_2)) &\to& f(q_{\text{id}}(x_1), q_{\text{id}}(x_2))\\
q_{\text{id}}(a) &\to& a
\end{array}
\]
It should be clear that the new transducer is equivalent to the original one.
Moreover, the new transducer uses no parameters whatsoever, therefore it is
depth proper.
Given a depth proper mtt, we can decide the LSHI property as follows.
We consider input trees which contain exactly one special marked input leaf
(it will be marked by a state $p$ of the look-ahead automaton, to act as a place-holder for
any input tree for which the look-ahead automaton arrives in state $p$).
For such input trees, the mtt produces output trees which only contain nested state calls
to the special input leaf. The original transducer has LSHI if and only if the
range of this transducer is finite (which is known to be decidable~\cite{DBLP:journals/iandc/DrewesE98}).
In a similar way we can decide LHI: here we consider input trees with multiple
marked input leaves.
To show that if such ranges are not finite, then the translation does not
have LSHI (or LHI), is done via pumping arguments (which use depth properness);
these pumping arguments are technically rather involved, but are (somewhat)
similar to the ones used in~\cite{DBLP:journals/siamcomp/EngelfrietM03}
to show that it is decidable whether or not an mtt has \emph{linear size increase} (LSI).

%In fact, it is known that
If we restrict the translations of mtts to LSI,
then we obtain exactly the MSO definable tree translations~\cite{DBLP:journals/siamcomp/EngelfrietM03}.
Note that this class of translation has recently been characterized by  new
models of tree transducers, the \emph{streaming tree transducer}~\cite{DBLP:journals/jacm/AlurD17} and
even more recently the \emph{register tree transducer}~\cite{DBLP:conf/lics/BojanczykD20}.
The LSI property is decidable for mtts
(it can even be decided for compositions of mtts, and if so,
then the translation is effectively MSO definable~\cite{DBLP:journals/acta/EngelfrietIM21}).
To decide LSI,
the given mtt is first transformed into ``proper'' normal form.
Properness guarantees that 
(1)~each state (except possibly the initial state) produces
infinitely many output trees 
%(this is called ``input proper''),
and that 
(2)~each parameter of a state is instantiated with infinitely many
distinct argument trees 
%(this is called ``parameter proper'').
Note that input properness is a 
generalization of the proper form of~\cite{DBLP:journals/iandc/AhoU71}.
Once in proper normal form, it suffices to check if the transducer
is ``finite copying''. This means that 
(a)~each node of each input tree is processed only a bounded number of times
and that
(b)~each parameter of every state is copied only a bounded number of times. 
\section{Preliminaries}

The set $\{0,1,\dots \}$ of natural numbers is denoted by $\mathbb{N}$.
For $k\in\mathbb{N}$ we denote by $[k]$ the set
$\{1,\dots,k\}$; thus $[0]=\emptyset$.
A ranked alphabet (set) consists of an alphabet (set) $\Sigma$ together
with a mapping $\text{rank}_{\Sigma}: \Sigma\to\mathbb{N}$
that assigns each symbol $\sigma\in\Sigma$ a natural number called its ``rank''.
We write $\sigma^{(k)}\in\Sigma$ to denote that $\sigma\in\Sigma$ and
$\text{rank}_{\Sigma}(\sigma)=k$.
By $\Sigma^{(k)}$ we denote the symbols of $\Sigma$ that have rank $k$.

The set $T_\Sigma$ of (finite, ranked, ordered) trees over $\Sigma$ is the smallest
set $S$ such that if $\sigma\in\Sigma^{(k)}$, $k\geq 0$, and
$s_1,\dots,s_k\in S$, then also $\sigma(s_1,\dots,s_k)\in S$.
We write $\sigma$ instead of $\sigma()$.
For a tree $s=\sigma(s_1,\dots,s_k)$ with $\sigma\in\Sigma^{(k)}$, $k\geq 0$,
and $s_1,\dots,s_k\in T_\Sigma$, we define 
the set $V(s)\subseteq\mathbb{N}^*$ of nodes of $s$ as
$\{\varepsilon\}\cup \{ iu\mid i\in[k], u\in V(s_i)\}$;
thus, nodes are strings over positive integers.
Thus, $\varepsilon$ denotes the root node of $s$, and
for a node $u$, $ui$ denotes the $i$-th child of $u$.
For $u\in V(s)$ we denote by $s[u]$ the label of $u$ in $s$ and
by $s/u$ the subtree rooted at $u$.
Formally, let $s=\sigma(s_1,\dots,s_k)$ and define 
$s[\epsilon]=\sigma$,
$s[iu]=s_i[u]$, 
$s/\varepsilon=s$, and
$s/iu=s_i/u$ for $\sigma\in\Sigma^{(k)}$, $k\geq 0$,
$s_1,\dots,s_k\in T_\Sigma$, $i\geq 1$ and $u\in V(s_i)$ such that $iu\in V(s)$.

We fix two special sets of symbols: the set $X=\{x_1,x_2,\dots \}$ of
variables and the set $Y=\{y_1,y_2,\dots\}$ of parameters.
For $k\geq 1$ let $X_k=\{x_1,\dots,x_k\}$ and $Y_k=\{y_1,\dots,y_k\}$.
Let $A$ be a set that is disjoint from $\Sigma$. Then the set $T_\Sigma(A)$ of
trees over $\Sigma$ indexed by $A$ is defined as $T_{\Sigma'}$ where
$\Sigma'=\Sigma\cup A$ and $\text{rank}_{\Sigma'}(a)=0$ for $a\in A$ and 
$\text{rank}_{\Sigma'}(\sigma)=\text{rank}_{\Sigma}$ for $\sigma\in\Sigma$.

For a ranked alphabet $\Sigma$ and a set $A$ the ranked set
$\<\Sigma,A>$  consists of all symbols $\<\sigma,a>$ with $\sigma\in\Sigma$ and
$a\in A$; the rank of $\<\sigma,a>$ is defined as $\text{rank}_\Sigma(\sigma)$.

\subsection{Tree Substitution}

Let $\Sigma$ be a ranked alphabet and let $s,t\in T_\Sigma$.
For $u\in V(s)$ we define the tree $s[u\leftarrow t]$ that is obtained
from $s$ by replacing the subtree rooted at node $u$ by the tree $t$.
Let $\sigma_1,\dots,\sigma_n\in\Sigma^{(0)}$, $n\geq 1$ be pairwise distinct symbols
and let $t_1,\dots,t_n\in T_\Sigma$. Then $t[\sigma_i\leftarrow t_i\mid i\in[n]]$
is the tree obtained from $t$ by replacing each occurrence of $\sigma_i$ by
the tree $t_i$. We have defined trees as particular strings, and this is just
ordinary string substitution (because we only replace symbols of rank zero).
We refer to this as ``first-order tree substitution''.

In ``second-order tree substitution'' it is possible to replace internal nodes $u$ (of a tree $s$)
by new trees. These new trees use parameters to indicate where
the ``dangling'' subtrees $s/ui$ of the node $u$ are to be placed.
Let $\sigma_1\in\Sigma^{(k_1)},\dots, \sigma_n\in\Sigma^{(k_n)}$ be pairwise distinct symbols 
with $n\geq 1$ and
$k_1,\dots,k_n\in\mathbb{N}$ and let $t_i\in T_\Sigma(Y_{k_i})$ for $i\in[n]$.
Let $s\in T_\Sigma$.
Then $s[\![\sigma_i\leftarrow t_i\mid i\in[n]]\!]$ denotes the tree
that is inductively defined as (abbreviating $[\![\sigma_i\leftarrow t_i\mid i\in[n]]\!]$ by
$[\![\dots ]\!]$) follows:
for $s=\sigma(s_1,\dots,s_k)$,
if $\sigma\not\in\{\sigma_1,\dots,\sigma_n\}$ then $s[\![\dots]\!]=\sigma(s_1[\![\dots]\!],
\dots,s_k[\![\dots]\!])$ and if $\sigma=\sigma_j$ for some $j\in[n]$ then
$s[\![\dots]\!]=t_j[y_i\leftarrow s_i[\![\dots]\!]\mid i\in[k]]$.

\subsection{Macro Tree Transducers}
\label{sect:mtt}

A (deterministic bottom-up) \emph{tree automaton} $A$ is given by a tuple
$(P,\Sigma,h)$ where $P$ is a finite set of states, $\Sigma$ is a ranked alphabet,
and $h$ is a collection of mappings $h_\sigma:P^k\to P$ with $\sigma\in\Sigma^{(k)}$
and $k\geq 0$ such that the collection is surjective (i.e., for every $p\in P$ there
exists a $\sigma$ such that $p$ is in the range of $h_\sigma$).
The extension of $h$ to a mapping $\hat{h}:T_\Sigma\to P$
is defined recursively as $\hat{h}(\sigma(s_1,\dots,s_k))=
h_\sigma(\hat{h}(s_1),\dots,\hat{h}(s_k))$ for every $\sigma\in\Sigma^{(k)}$,
$k\geq 0$, and $s_1,\dots,s_k$. For every $p\in P$ we define the subset $L_p$ of trees
in $T_\Sigma$ as $\{ s\in T_\Sigma\mid \hat{h}(s)=p\}$.
Note that $L_p\not=\emptyset$ due to the surjectivity requirement above.

A (total, deterministic) \emph{macro tree transducer with (regular) look-ahead} (``mttr'') 
$M$ is given by a tuple
$(Q,P,\Sigma,\Delta,q_0,R,h)$, where
\begin{itemize}
\item $Q$ is a ranked alphabet of \emph{states},
\item $\Sigma$ and $\Delta$ are ranked alphabet of \emph{input} and \emph{output symbols},
\item $(P,\Sigma,h)$ is a tree automaton (called the \emph{look-ahead automaton} of $M$),
\item $q_0\in Q^{(0)}$ is the \emph{initial state},
\item and $R$ is the \emph{set of rules}, where for each $q\in Q^{(m)}$, $m\geq 0$,
$\sigma\in\Sigma^{(k)}, k\geq 0$, and $p_1,\dots, p_k\in P$ there is exactly one rule of the form 
\[
\<q,\sigma(x_1:p_1,\dots,x_k:p_k)>(y_1,\dots,y_m) \to t
\]
with $t\in T_{\Delta\cup \< Q,X_k>}(Y_m)$.

The right-hand side $t$ of such a rule is denoted by
$\text{rhs}_M(q,\sigma,\<p_1, \dots,p_k>)$
\end{itemize}

We use a notation that is slightly different from the one used in the Introduction:
instead of, e.g., $q_2(x_2, q_{\text{id}}(x_1))$ we write
$\< q_2,x_2>(\< q_{\text{id}},x_1>)$.
Thus, we use angular brackets $\< \dots >$ to indicate a state call on an input subtree,
and use round brackets (after the angular brackets), to indicate the parameter arguments
of the particular state call.

The semantics of an mttr $M$ (as above) is defined as follows.
We define the derivation relation $\Rightarrow_M$ as follows.
For two trees $\xi_1,\xi_2\in T_{\Delta\cup\< Q,T_\Sigma>}(Y)$,
$\xi_1\Rightarrow_M\xi_2$ if there exists a node $u$ in $\xi_1$
with $\xi_1/u=\<q,s>(t_1,\dots,t_m)$, $q\in Q^{(m)}$, $m\geq 0$,
$s=\sigma(s_1,\dots,s_k)$, $\sigma\in\Sigma^{(k)}$, $k\geq 0$,
$s_1,\dots,s_k\in T_\Sigma$,
$t_1,\dots,t_m\in T_{\Delta\cup\<Q,T_\Sigma>}(Y)$, and 
$\xi_2=\xi_1[u\leftarrow\xi]$ where $\xi$ equals
\[
\zeta
[\![\<q',x_i>\leftarrow \<q',s_i>\mid q'\in Q,i\in[k]]\!]
[y_j\leftarrow t_j\mid j\in[m]]
\]
and $\zeta=\text{rhs}_M(q,\sigma,\<\hat{h}(s_1),\dots,\hat{h}(s_k)>)$.
Since $M$ is total (i.e.,
for every state $q$, input symbol $\sigma\in\Sigma^{(k)}$, $k\geq 0$, and 
look-ahead states $p_1,\dots,p_k$, $M$ contains a corresponding rule)
%and $\Rightarrow_M$ is confluent and terminating,
there is for every $\xi_1$ a unique
tree $\xi'\in T_\Delta(Y)$ such that $\xi_1\Rightarrow_M^* \xi'$.
For every $q\in Q^{(m)}$, $m\geq 0$ and $s\in T_\Sigma$ we define the
\emph{$q$-translation of $s$}, denoted by $M_q(s)$, as the unique tree $t$ in $T_\Delta(Y_m)$ such that
$\<q,s>(y_1,\dots, y_m)\Rightarrow_M^* t$.
We denote the \emph{translation realized by $M$} also by $M$, i.e., 
$M=M_{q_0}$ and for every $s\in T_\Sigma$,
$M(s)=M_{q_0}(s)$ is the unique tree $t\in T_\Delta$
such that $\< q_0,s>\Rightarrow_M^* t$.

Let $M$ be an mttr as before.
We define \emph{the extension $\widehat{M}$ of $M$} which can also process
look-ahead states at leaves of input trees.
Let $\widehat{M}=(Q,P,\hat{\Sigma},\hat{\Delta},q_0,R\cup \hat{R},h\cup h')$ where
$\hat{\Sigma}=\Sigma\cup \{p^{(0)}\mid p\in P\}$ and
$\hat{\Delta}=\Delta\cup \{\< q,p>^{(m)} \mid q \in Q^{(m)}, p \in P, m\geq 0\}$.
For every $q\in Q^{(m)}$, $m\geq 0$, and $p\in P$ we let $h(p)=p$ and we
let the rule
$\<q,p>(y_1, \dots, y_m)\to \<q,p>(y_1, \dots, y_m)$ be in $\hat{R}$;
note that the $\<q,p>$ on the right-hand side of this rule is an output symbol.
For the original transducer $M$ we say that the pair $(q,p)$ is
\emph{reachable (in $M$)} if there is an input tree $s\in T_{\hat{\Sigma}}$ such
that $\<q,p>$ occurs in $\widehat{M}(s)$.
Clearly it is decidable for a given pair $(q,p)$, whether or not
it is reachable; this is because (1)~inverse translations of mttrs
effectively preserve regularity~\cite{DBLP:journals/jcss/EngelfrietV85,DBLP:journals/ipl/PerstS04},
(2)~the set of all trees in $T_{\hat{\Delta}}$
that contain at least one occurrence of $\< q,p>$ is regular, and
(3)~emptiness of regular tree languages is decidable~\cite{TATA07}. 

We say that $M$ is \emph{nondeleting}, if for every state $q\in Q^{(m)}$,
$m\geq 1$, $\sigma\in\Sigma^{(k)}, k\geq 0$, $p_1,\dots, p_k\in P$,
and $j\in[m]$, there is at least one occurrence of $y_j$ in
$\text{rhs}_M(q,\sigma,\< p_1,\dots,p_k>)$.
The next proposition is proved in~\cite[Lemma~6.7]{DBLP:journals/iandc/EngelfrietM99}
(for mttrs that do not copy parameters, but the proof works analogously for
arbitrary mtts).

\begin{proposition}
\label{prop:nondeleting}
For every mttr, an equivalent nondeleting mttr $M$ can be constructed such that,
for every state $q$ of $M$ of rank $m\geq 1$ and
for every $j\in[m]$ it holds that $M_q(s)$ contains at least one
occurrence of $y_j$, for every $s\in T_\Sigma$.
\end{proposition}

It is well known that the finiteness of ranges of compositions of 
mttrs is decidable~\cite{DBLP:journals/iandc/DrewesE98}.
A (partial nondeterministic) top-down tree transducer with look-ahead
(``topr'' for short)
is an mttr
as before, where $Q=Q^{(0)}$ and $R$ may contain none or several
rules for each given $q$ and $\sigma$.

\begin{proposition}(\cite[Theorem~4.5]{DBLP:journals/iandc/DrewesE98})
\label{prop:finite}
For a given composition of mttrs and (partial nondeterministic) toprs it is 
decidable whether or not the range of the composition is finite. 
In the case of finiteness, the range can be constructed.
\end{proposition}

%Note that a ``partial'' mttr is one where for some $q,\sigma,p_1,\dots,p_k$
%it may be that $\text{rhs}_M(q,\sigma,\< p_1,\dots,p_k>)$ is \emph{not}
%defined.

%!TEX root = main.tex
\section{Depth Proper Normal Form}

The depth proper normal form requires that each parameter of
each state $q$ occurs at unbounded depth in the output trees of that state 
(for each given look-ahead state $p$ such that $(q,p)$ is reachable).
Formally, let $q$ be a state of rank $m\geq 1$, $j\in[m]$, and
$p\in P$. 
If $(q,p)$ is reachable, then 
for every natural number $n$ there must exist an input tree $s_n \in L_p$ such that
$y_j$ occurs at depth $>n$ in the tree $M_q(s_n)$.
Conversely, we say that parameter $y_j$ is \emph{depth-bounded} for $q$ and $p$ 
if there exists an $n$ for which no such input tree $s_n \in L_p$ exists;
more generally, we say that $Z\subseteq Y_m$ is
\emph{depth-bounded} for $q$ and $p$, if each $y\in Z$ is depth-bounded for $q$ and $p$.

If $Z$ is depth-bounded for $q$ and $p$, then there are only finitely many output paths
in the trees in $M_q(L_p)$ under which the parameters from $Z$ occur.
The \emph{$Z$-skeleton} of an arbitrary tree $t$ is obtained from $t$
by replacing each top-most node $u$ such that $t/u$ does not contain any
occurrence of a parameter from $Z$ by some symbol. 
Clearly, $Z$ is depth-bounded for $q$ and $p$ if and only if the $Z$-skeleta of
all trees in $M_q(L_p)$ form a finite set.

Let $\Delta$ be an arbitrary ranked alphabet, $m\geq 1$, 
$t\in T_\Delta(Y_m)$, and $Z\subseteq Y_m$.
Let us write \(\paras{t}\subseteq{Y_m}\) for the set of parameters 
occurring in $t$.
Let us now be more specific as to which symbols replace the top-most nodes $u$
of $t$ such that $\paras{t/u}\cap Z=\emptyset$. Since in our construction later
we will want to obtain a transducer that is nondeleting, it will be helpful
to know which parameters appear in a given deleted tree. Therefore we replace
such nodes $u$ by the set $\paras{t/u}$. 
We denote by $\lcop{t}_Z$ the $Z$-skeleton of $t$ and define it inductively as follows
(where $\delta\in\Delta$):
\begin{align*}
\lcop{t}_Z &= 
\begin{cases}
t &
\text{if \(t\in Z\)}
\\
\delta(\lcop{t_1}_Z,\dots,\lcop{t_n}_Z) &
\text{if \(\paras{t}\cap Z\ne\emptyset\) and \(t = \delta(t_1,\dots,t_n)\)}
\\
\paras{t} & \text{if \(\paras{t}\cap Z = \emptyset\)}.
\end{cases}
\end{align*}

The definition of $\lcop{t}_Z$ is extended to sets $L$ of trees
as $\lcop{L}_Z=\{\lcop{t}_Z\mid t\in L\}$.
We call \emph{$Y$-nodes} the nodes $u$ in $V(\lcop{t}_Z)$ such 
that $\lcop{t}_Z/u=Z'\subseteq Y_m$. We denote by 
$\mathcal{U}(\lcop{t}_Z)$ the set of $Y$-nodes on $\lcop{t}_Z$. 
The notion of parameters in a tree naturally extends to $Z$-skeleta with, 
for a $Y$-node labeled $Z'$: $\paras{Z'} = Z'$. 
The proof of the next lemma is straightforward by induction on $t$
(see the Appendix).

\begin{lemma}\label{lm:nd}
Let $\Delta$ be a ranked alphabet, $m\geq 1$, $Z\subseteq Y_m$, and
$t\in T_\Delta(Y_m)$.
(1)~$t=\lcop{t}_Z[u\leftarrow t/u\mid u\in\mathcal{U}(\lcop{t}_Z)]$.
(2)~$\paras{\lcop{t}_Z} = \paras{t}$. 
%(3)~If $y\in Y_m$ occurs in $t$, then 
%there exists a leaf $u\in V(\lcop{t}_Z)$
%such that either $\lcop{t}_Z/u=y$ or 
%$\lcop{t}_Z/u=Z'\subseteq Y_m$ and $y\in Z'$.
\end{lemma}

%%Sebastian's version:
%The proof is by induction on the structure of $t$.
%Let us denote the substitution $[u\leftarrow t/u\mid u\in\mathcal{U}(\lcop{t}_Z)]$
%by $[t]$.
%%
%We first consider $t=y\in Y$. There are two cases.
%
%Case 1: $y\in Z$. Then $\mathcal{U}(\lcop{t}_Z)=\emptyset$.
%Hence $\lcop{t}_Z[t]=\lcop{y}_Z$. The latter equals $y=t$ by 
%the definition of $\lcop{.}_Z$. Thus~(1)~holds and~(2)~holds for $u=\epsilon$.
%
%Case 2: $y\not\in Z$. Then $\mathcal{U}(\lcop{t}_Z)=\{ \epsilon \}$ and hence
%$\lcop{t}_Z[t] = \lcop{t}_Z[\epsilon\leftarrow t/\epsilon = t] = t$ which proves~(1).
%Moreover, $\lcop{t}_Z=\{y_j\}$ by the definition of $\lcop{.}_Z$,
%i.e., again~(2)~holds for $u=\varepsilon$.
%
%Now consider $t=\delta(t_1,\dots,t_n)$, $\delta\in\Delta^{(n)}$, $n\geq 0$,
%and $t_1,\dots, t_n\in T_\Delta(Y_m)$.
%To show~(1) we obtain from the definition $\lcop{t}_Z$ that
%\[
%\lcop{t}_Z[t]=\delta(\lcop{t_1}_Z,\dots,\lcop{t_n}_Z)[t]
%= \delta(\lcop{t_1}_Z[t_1],\dots,\lcop{t_n}_Z[t_n]),
%\]
%where for $i\in[n]$, $[t_i]$ denotes the substitution
%$[u\leftarrow t_i/u\mid u\in\mathcal{U}(\lcop{t_i}_Z)]$.
%By induction the latter equals $\delta(t_1,\dots,t_n)=t$.
%To show~(2) we assume that $y_j$ occurs in $t$ for some $j\in[m]$.
%Thus, $n\geq 1$ and there exists an $i\in[n]$
%such that $y_j$ occurs in $t_i$.
%By the induction for Statement~(2), there exists a leaf $u$ of $t_i$ such that 
%the Statement~(2) holds (for $u$ and $t_i$), and therefore Statement~(2) holds
%for $iu$ and $t$.
%\end{proof}

Finally, we define depth properness for mtts with look-ahead. 

\begin{definition} \label{df_depth_proper}
The mttr $M=(Q,P,\Sigma,\Delta,q_0,R,h)$ is in \emph{depth proper normal form}
(or, synonymously, $M$ is \emph{depth proper})
if for every $q\in Q^{(m)}$, $m\geq 1$, and $p\in P$ it holds that
if $(q,p)$ is reachable, then  
$\lcop{M_q(L_p)}_{\{y_j\}}$ is infinite for all $j\in[m]$.
\end{definition}

From now on we will want to make use of the following definitions:
\[
\begin{array}{lcl}
F_p &=& \{ q\in Q^{(m)}\mid \exists j\in[m], \lcop{M_q(L_p)}_{\{y_j\}}
\text{ is finite}\}\\
Y(q,p) &=& \{ y_j\mid j\in[\text{rank}_Q(q)]\text{ such that }
\lcop{M_q(L_p)}_{\{y_j\}}\text{ is finite}\}
\end{array}
\]

It should be clear that $\lcop{M_q(L_p)}_{Y(q,p)}$ is finite for every $q$ and $p$,
as stated in the next lemma (the proof is in the Appendix).

\begin{lemma}\label{lm:lcop_finite}
Let $M$ be an mtt, $q$ a state of $M$, and $p$ a look-ahead state of $M$
such that $Y(q,p)\not=\emptyset$.
Then $\lcop{M_q(L_p)}_{Y(q,p)}$ is finite.
\end{lemma}

\subsection{Construction of the Normal Form and Examples}
\label{sect:examples}

%\subsection{Formal Construction of the Depth Proper Normal Form}
%\label{sect:dp}

Let $M$ be an mttr as before. 
We assume that $M$ is nondeleting (which is justified by 
Proposition~\ref{prop:nondeleting}).
The idea of the construction is as follows.
First, we determine all reachable pairs $(q,p)$ 
such that $Y(q,p)\not=\emptyset$.
Let $(q,p)$ be such a pair and let $Z=Y(q,p)$.
An occurrence of $\<q,x_i>$ in a right-hand side
$\text{rhs}_M(q',\sigma,\< p_1,\dots,p_k>)$ such that $p_i=p$ is
called a \emph{$(q,p)$-call}. Our aim is to replace each $(q,p)$-call
by an appropriate tree from $\lcop{M_q(L_p)}_Z$. Just which tree is
the appropriate one will be determined by regular look-ahead. 
Moreover, such trees should be modified not to contain leaf nodes 
labeled by subsets of $Y$: such nodes will be replaced by 
calls of new ``helper states''. 

%We now present the formal construction of the transducer $\pi(M)$.
%Note that in general, $\pi(M)$ is \emph{not} depth proper yet, but
%the construction needs to be iterated several times (cf. the examples in
%Section~\ref{sect:examples}).

\begin{definition}
\label{df:pi}
Let $M=(Q,P,\Sigma,\Delta,q_0,R,h)$ be a nondeleting mttr
that is not depth proper.
We construct the new mttr
$\pi(M)=(Q\cup H,P',\Sigma,\Delta,q_0,R',h')$.
% The sets $H$ and $P'$ are defined as follows.
For every 
$q\in Q^{(m)}$, $m\geq 1$, and $p\in P$ such that $Y(q,p)\not=\emptyset$,
$H$ contains the following set of helper states:
\[
\{ [q,p,t,u]^{(|U|)}\mid 
t\in \lcop{M_q(L_p)}_{Y(q,p)}, 
u\in V(t),
t/u=U\subseteq Y_m \}
\]
and $P'$ contains $(p,\varphi)$ for any
function $\varphi$ that assigns to each $q\in F_p$ 
a tree in $\lcop{M_q(L_p)}_{Y(q,p)}$.
Observe that $H$ and $P'$ are well defined, because
$\lcop{M_q(L_p)}_{Y(q,p)}$ is finite by Lemma~\ref{lm:lcop_finite}. 
Note that $|U|\leq |Y_m \setminus Y(q,p)|$; since $Y(q,p)$ is non-empty this implies that
the rank of each helper state is at most $(r-1)$, where $r$ is the maximal
rank of the states in $Q$. 

For every $q\in Q^{(m)}$, $m\geq 0$, 
$\sigma\in\Sigma^{(k)}$, $k\geq 0$, and 
$(p_1,\varphi_1),\dots,(p_k,\varphi_k)\in P'$ we let the rule
\[
\<q,\sigma(x_1:(p_1,\varphi_1),\dots,x_k:(p_i,\varphi_i)>(y_1,\dots,y_m)
\to\text{rhs}_M(q,\sigma,\<p_1,\dots,p_k>)[\![ . ]\!]
\]
be in $R'$, where the second-order tree substitution
$[\![ . ]\!]$ is defined as follows.
\begin{multline*}
[\![ . ]\!] = 
[\![\< q',x_i>\leftarrow \varphi_i(q')\big[
u\leftarrow [q',p_i,\varphi_i(q'),u](y_{j_1},\dots,y_{j_n})\mid \\
\varphi_i(q')/u=\{y_{j_1},\dots,y_{j_n}\},
j_1<\cdots <j_n\big] \mid q'\in F_{p_i}, i\in[k]
 ]\!].
\end{multline*}
We define $h_\sigma'((p_1,\varphi_1),\dots,(p_k,\varphi_k))=(p,\varphi)$
where $p=h_\sigma(p_1,\dots,p_k)$ and, using special second-order substitution 
from Definition~\ref{def:meta-skeleta}, for every $q\in F_p$,  
\[
\varphi(q)=
\lcop{
\text{rhs}_M(q,\sigma,\<p_1,\dots,p_k>)
[\![\< q',x_i>\leftarrow \varphi_i(q')\mid q'\in F_{p_i}, i\in[k]]\!]\su
}_{Y(q,p)}.
\]
For every helper state $[q,p,t,u]\in H^{(n)}$, $n\geq 0$,
$\sigma\in\Sigma^{(k)}$, $k\geq 0$, and 
$(p_1,\varphi_1),\dots,$ $(p_k,\varphi_k)\in P'$ 
such that $h_\sigma(p_1,\dots,p_k)=p$
we let the rule
\[
\<[q,p,t,u](\sigma(x_1:(p_1,\varphi_1),\dots,x_k:(p_k,\varphi_k))>(y_1,\dots,y_n)
\to \xi/u[y_{j_\nu}\leftarrow y_\nu\mid\nu\in[n]]
\]
be in $R'$ where
$t/u=\{y_{j_1},\dots,y_{j_n}\}$,
$j_1<\cdots <j_n$,
$\xi=\text{rhs}_M(q,\sigma,\<p_1,\dots,p_k>)[\![ . ]\!]$, and
$[\![ . ]\!]$ is the substitution from above.
\end{definition}

%We present two examples to show how a depth proper mttr is obtained.

\newcommand\qid{q_{\mathrm{id}}}

We now show how the depth proper normal form is achieved using
an example. An additional example (which makes more interesting use of
helper states) can be found in the Appendix. 
%
%Next we shall show a more complex example where
%a set of \(Z\)-skeleta is not singleton and 
%a $Y$-node of a \(Z\)-skeleton is not empty
%(and where the construction needs to be iterated).
%
Let \(M=(Q,\{p\},\Sigma,\Delta,q_0,R,h_0)\) 
with
\(Q=\{q_0^{(0)},q_1^{(1)},q_2^{(2)}\}\),
\(\Sigma=\{a^{(1)},b^{(1)},e^{(0)}\}\),
and \(\Delta=\{f^{(2)},g^{(1)},e^{(0)}\}\)
be an mttr 
where \((\Sigma,\{p\},h_0)\) with $L_p=T_\Sigma$
and \(R\) consists of these rules:
\begin{align*}
\<q_1, a(x)>(y_1) &\to \<q_2, x>(y_1, \<q_1,x>(y_1)) &
\<q_2, a(x)>(y_1,y_2) &\to f(y_1,\<q_1,x>(g(y_2)))
\\
\<q_1, b(x)>(y_1) &\to y_1 &
\<q_2, b(x)>(y_1,y_2) &\to f(y_2,y_1)
\\
\<q_1, e>(y_1) &\to g(y_1) &
\<q_2, e>(y_1,y_2) &\to f(y_2,y_1)
\end{align*}
We suppose that the \(q_0\)-rules are defined so that all states are reachable.
Now we have \(F_p=\{q_2\}\), \(Y(q_2,p)=\{y_1\}\), and 
\(\lcop{M_{q_2}(L_p)}_{\{y_1\}} = \{ t_1, t_2 \}\)
with \(t_1 = f(y_1,\{y_2\})\) and \(t_2=f(\{y_2\},y_1)\).
As before, we can rewrite \(q_2\)-calls with the skeleta,
but since there are two possibilities \(t_1\) and \(t_2\), 
we need to separate the rules according to the input using the look-ahead.
In general, \(F_p\) contains several states each of which may have multiple skeleta, 
so each look-ahead contains a finite map from \(F_p\) to skeleta.
Let \(\varphi_1=\{q_2\mapsto t_1\}\) and \(\varphi_2=\{q_2\mapsto t_2\}\) 
such that \(L_{p,\varphi_1}=\{a(s)\mid s\in\tree\Sigma\}\) and
\(L_{p,\varphi_2}=\{b(s)\mid s\in\tree\Sigma\}\cup\{e\}\).
The \((q_1,a)\)-rule containing a \(q_2\)-call is separated as
\begin{align*}
\<q_1,a(x:(p,\varphi_1))>(y_1) &\to f(y_1, \<[q_2,p,t_1,2],x>(\<q_1,x>(y_1))) \\
\<q_1,a(x:(p,\varphi_2))>(y_1) &\to f(\<[q_2,p,t_2,1],x>(\<q_1,x>(y_1)),y_1) 
\end{align*}
where \(\<[q_2,p,t_1,2],x>\) and \(\<[q_2,p,t_2,1],x>\) are helper states.
Each helper state has rank 1 
because the corresponding node in the skeleton is a $Y$-node of length 1.
The arguments of the call are inherited from
the arguments of the original \(q_2\)-call that occur in the sequence.
For example, \(\<[q_2,p,t_1,2],x>\) is called with \(\<q_1,x>(y_1)\)
since \(t_1\) has a $Y$-node \(\{y_2\}\) and 
the original \(q_2\)-call has \(\<q_1,x>(y_1)\) as the second argument.
The rules of these helper states are constructed from the original \(q_2\)-rule
with substitution (which causes nothing since no states in \(F_p\) are called)
and extracting a subtree at the $Y$-node, that is,
\[
\begin{array}{lcl}
\<[q_2,p,t_1,2], a(x:(p,\varphi))>(y_1) &\to& \<q_1,x>(g(y_1)) \\
\<[q_2,p,t_2,1], b(x:(p,\varphi))>(y_1) &\to& y_1 \\
\<[q_2,p,t_2,1], e>(y_1) &\to& y_1 
\end{array}
\]
where \(\varphi\in\{\varphi_1,\varphi_2\}\) and
we had to rename the parameter \(y_2\) into \(y_1\)
(because the helper states only refer to \(y_2\)).
Note that rules for \(([q_2,p,t_1,2],b)\), \(([q_2,p,t_1,2],e)\) and \(([q_2,p,t_2,1], a)\) 
do not have to be considered.
These rules are not referred because
the states are never called with the input symbols
due to their look-ahead.
For example, the \([q_2,p,t_1,2]\)-call occurs only 
in the \((q_1,a)\)-rule with \(x\in L_{p,\varphi_1}\)
in which the root symbol cannot be \(b\).
%They are not reachable because they are called with the specific look-ahead.

Thereby we have been able to remove every call of states in \(F_p\).
However, new improper states may be generated by the separation of rules 
because of the look-ahead introduction.
In fact, we have \(F_{p,\varphi_2}=\{q_1,q_2,[q_2,p,t_2,1]\}\) in the example above.
Since every \(q_2\)-call has already been removed in the previous step,
we have to apply the same technique again for the calls of 
\(q_1\) and \([q_2,p,t_2,1]\).
We have $Y(q_1,(p,\varphi_2))=\{y_1\}$ and 
$Y([q_2,p,t_2,1],(p,\varphi_2))=\{y_1\}$.
Moreover 
$\lcop{M'_{q_1}(L_{p,\varphi_2})}_{\{y_1\}}=\{y_1, g(y_1)\}$
and
$\lcop{M'_{[q_2,p,t_2,1]}(L_{p,\varphi_2})}_{\{y_1\}}=\{y_1\}$.

\iffalse
We have: \vspace{-2mm}
\begin{align*}
&Y(q_1,(p,\varphi_2))=\{y_1\}
&&\text{and}
&Y([q_2,p,t_2,1],(p,\varphi_2))=\{y_1\}, \\
&\lcop{M'_{q_1}(L_{p,\varphi_2})}_{\{y_1\}}=\{y_1, g(y_1)\}
&&\text{and}
&\lcop{M'_{[q_2,p,t_2,1]}(L_{p,\varphi_2})}_{\{y_1\}}=\{y_1\}\text.
\end{align*} \vspace{-5mm}
\fi

Look-ahead has to be introduced to determine which skeleton to output.
Two maps over \(F_{p,\varphi_2}\),
except for \(q_2\) whose call has already been removed,
are defined: 
\(\varphi_3=\{q_1\mapsto y_1, [q_2,p,t_2,1]\mapsto y_1\}\) and
\(\varphi_4=\{q_1\mapsto g(y_1), [q_2,p,t_2,1]\mapsto y_1\}\)
such that \(L_{p,\varphi_2,\varphi_3}=\{b(s)\in L_{p,\varphi_2}\mid s\in\tree\Sigma\}\) and
\(L_{p,\varphi_2,\varphi_3}=\{e\}\).
The \((q_1,a)\)- and \(([q_2,p,t_1,2],a)\)-rules with look-ahead \(\varphi_2\)
which contains a \(q_2\)-call 
are separated as follows:
\[
\begin{array}{lcl}
\<q_1, a(x:(p,\varphi_2,\varphi_3))>(y_1) &\!\to& f(y_1,y_1) \\
  \<[q_2,p,t_1,2], a(x:(p,\varphi_2,\varphi_3))>(y_1) &\!\to& g(y_1) \\
\<q_1, a(x:(p,\varphi_2,\varphi_4))>(y_1) &\!\to& f(g(y_1),y_1) \\
  \<[q_2,p,t_1,2], a(x:(p,\varphi_2,\varphi_4))>(y_1) &\!\to& g(g(y_1)) 
\end{array}
\]
The resulting  mttr is depth proper.

\subsection{Correctness Proof and Termination of Iteration}
\label{sect:corr}

Here we prove the correctness of transducer $\pi(M)$ that was defined
in Definition~\ref{df_depth_proper}. Lemma~\ref{lm:corr} establishes
the correctness of the look-ahead, relates the states of $\pi(M)$ to those
of $M$, and shows that the transducer $\pi(M)$ is nondeleting.
The latter is needed, so that the construction of $\pi$ can be carried
out iteratively (recall from Definition~\ref{df_depth_proper} that $M$ is
required to be nondeleting, in order to construct $\pi(M)$).
The proof of the following lemma is somewhat involved but
rather technical and can be found in the Appendix; to prove Point~(2)
it uses a ``special'' kind of second-order tree substitution
which replace $Y$-nodes by new Y-nodes consisting of parameters in the output trees that
would have been substituted for the parameters in the original $Y$-node.

\begin{lemma}
\label{lm:corr}
Let $M$ be a nondeleting mttr and $N=\pi(M)$ be the mttr of
Definition~\ref{df:pi}, both with the tuples as in that definition. 
Let $s\in T_\Sigma$ with $\hat{h'}(s)=(p,\varphi)$.
%Then
\begin{enumerate}
\item[(1)] $p=\hat{h}(s)$,
\item[(2)] $\forall q\in F_p$: $\varphi(q)=\lcop{M_q(s)}_{Y(q,p)}$, 
\item[(3)] $\forall q\in Q$: $N_q(s)=M_q(s)$,
\item[(4)] $\forall q\in F_p$ and $u\in V(t)$ with $t=\varphi(q)$ and
$t/u=\{y_{j_1},\dots,y_{j_n}\}$ with \\
$j_1<\cdots <j_n$:
$N_{[q,p,t,u]}(s)=M_q(s)/u[y_{j_\nu}\leftarrow y_\nu\mid\nu\in[n]]$, and
\item[(5)] the mttr $N$ is nondeleting.
\end{enumerate}
\end{lemma}

%!TEX root = ICALP_main.tex

% macros (will be moved later)

%\section{Termination of the procedure}

We show that the iteration of the construction $\pi(M)$ will
terminate with a transducer that is depth proper.
%For an arbitrary transducer $M$ and a look-ahead state $p$ of $M$ we define
%the set $U_p$ of states $q$ such that $(q,p)$ us unreachable as follows''
%\[
%U_p = \{ q\in Q\mid (q,p)\text{ is unreachable in }M\}.
%\]
%Note that $M$ is depth proper if $F_p\subseteq U_p$ for every look-ahead state $p$ of $M$.
%
First, let us discuss what property a single iteration of $\pi$ ensures.
Let $p\in P$.
Note that the set $F_p$ is defined independently of reachability, i.e.,
$F_p$ may contain states $q$ such that $(q,p)$ is \emph{not} reachable.
Let $\varphi$ such that $(p,\varphi)\in P'$.
Then
%\[
$F_p\subseteq F_{(p,\varphi)}$.
%\]
This inclusion follows from Lemma~\ref{lm:corr} as follows:
let $s\in L_{(p,\varphi)}$ and let $q\in F_p$ be of rank $m$.
The latter means that there exists a $j\in[m]$ and a number $n$ 
such that every occurrence of $y_j$ in $M_q(s')$ is
at depth $\leq n$ for every $s'\in L_p$.
By Lemma~\ref{lm:corr}~(1), $s\in L_p$ and by Lemma~\ref{lm:corr}~(3),
$N_q(s)=M_q(s)$. Thus, every occurrence of $y_j$ in $N_q(s)$ also
occurs at depth $\leq n$. Hence, $q\in F_{(p,\varphi)}$.

We now consider reachability.
We say that a state $q$ is \emph{depth proper}, if for all $p\in P$ such that
$(q,p)$ is reachable, $q\not\in F_p$.
If $q\in F_p$, then for all $\varphi$ such that $(p,\varphi)\in P'$ it holds that
$(q,(p,\varphi))$ is not reachable.
This property follows immediately from the definition of look-ahead and the rules of $\pi(M)$:
the substitution $[\![.]\!]$ replaces each state call $\<q',x_i>$ with $q\in F_{p_i}$ by
a tree that does not contain states of $Q$.
So, if $(q,(p,\varphi))$ is reachable, then $q\not\in F_p$;
however, it may be that $q\in F_{(p,\varphi)}$, which means that $q$ is
not depth proper. It means that if $F_{(p,\varphi)}=F_p$ for
all $(p,\varphi)\in P'$, then all states $q\in Q$ are depth proper.
Let $Q_0=Q$ and consider now the iterated application of $\pi$.
Clearly, after some iterations of $\pi$, it will hold that
$F_{(p,\varphi)}=F_p$ for all $(p,\varphi)\in P'$.
To see this, consider the chain of inclusions
\[
F_p\cap Q_0 \subseteq F_{(p,\varphi_1)}\cap Q_0 \subseteq \dots \subseteq
   F_{(p,\varphi_1,\dots,\varphi_k)}\cap Q_0 \subseteq \dots
\]
for any maps \(\varphi_i\) introduced in the look-ahead of \(\pi^i(M)\).
Since \(Q_0\) is finite, the chain contains only finitely many strict inclusions.
Hence there is a minimal \(n\) such that 
\( F_{(p,\varphi_1,\dots,\varphi_n)}\cap Q_0 = F_{(p,\varphi_1,\dots,\varphi_{n'})} \cap Q_0 \)
for any \(n'>n\).

Consider a tree with an artificial root node which contains all such chains, i.e.,
for each $p\in P$ there is exactly one child of the root node labeled $F_p$, and 
a node labeled $F_p$ has children labeled $F_{(p,\varphi)}$ for each $(p,\varphi)\in P'$, etc. 
Moreover, a node labeled $F_{(p,\varphi_1,\dots,\varphi_n)}$ as in the chain above
is a leaf of this tree.
Since each node of this tree is finitely branching (because $P'$ is finite) and each
path has finite length, we know by K{\"o}nig's lemma that tree is finite.
Thus, if $d$ is the depth of this tree, for the 
mttr $M'=\pi^d(M)$, all states in $Q_0$ are depth proper.

Let $m$ be the maximal rank of the states in $Q_0$.
Since all helper states are of rank $<m$, we know that
$M'$ contains no improper states of rank $\geq m$.
We now proceed in the same fashion and construct a transducer $M''=\pi^{n'}(M')$
which contains no improper states of rank $\geq (m-1)$.
In a similar way we 
eventually obtain an mttr for which \emph{all states} are depth proper
(and which is equivalent to $M$).
Thus, even though we do not constructively derive a precise bound, we know that
after \emph{some} number of applications of $\pi$ we are sure to obtain
a depth proper mttr.

Before we state the main theorem of this section, we need the following
lemma.
%(the proof is a straightforward reduction to Proposition~\ref{prop:finite} and
%can be found in the Appendix).

\begin{lemma}\label{lm:dec}
Let $M=(Q,P,\Sigma,\Delta,q_0,R,h)$ be an mttr and let
$q\in Q^{(m)}$, $m\geq 1$, $j\in[m]$, and $p\in P$.
It is decidable whether or not 
$\lcop{M_q(L_p)}_{\{y_j\}}$ is finite.
In case of finiteness, 
$\lcop{M_q(L_p)}_{\{y_j\}}$ can be constructed.
\end{lemma}

Since for a pair $(q,p)$ it is decidable whether or not it is reachable
(see Section~\ref{sect:mtt}), Lemma~\ref{lm:dec} implies that
it is decidable whether or not a given mttr is depth proper.

\begin{theorem}
  \label{th:proper}
For every mttr \(M\), we can construct an equivalent mttr $M'$ such that
$M'$ is depth proper.
\end{theorem}
\begin{proof}
There is a nondeleting mttr \(M_0\) equivalent to \(M\) 
(\cite{DBLP:journals/iandc/EngelfrietM99} or Proposition~\ref{prop:nondeleting}).
We repeatedly construct equivalent transducers $\pi(M)$, $\pi(\pi(M))$, etc.
until a proper mttr is obtained (which is decidable by
Lemma~\ref{lm:dec}). The repetition terminates 
(first eliminating all reachable calls of improper states of the highest rank $m$, then
those or rank $m-1$, etc.).
%\qed
\end{proof}

%!TEX root = ICALP_main.tex
\section{Linear Height and Linear Size-to-Height Increase}
\label{sec:decision_LSHI}

Let $\Gamma$ be a ranked alphabet and $t$ a tree over $\Gamma$.
We define the size $|t|$ of a tree as its number of nodes $|V(t)|$.
The height $\he{t}$ of $t$ is defined as
$\he{t}=0$ if $t\in\Gamma^{(0)}$ and 
$\he{t} = 1 + \text{max}\{\he{t_i}\mid i\in[k]\}$ if
$t=\gamma(t_1,\dots,t_k)$ for $\gamma\in\Gamma^{(k)}$, $k\geq 1$,
and $t_1,\dots,t_k\in T_\Gamma$.

Let $M$ be an mttr (with input ranked alphabet $\Sigma$).
Then $M$ has \emph{linear size-to-height increase} (for short LSHI) if
there exists a number $c$
such that for every input tree
$s\in T_\Sigma$: $\he{M(s)}\leq c\cdot |s|$.
The mttr $M$ has \emph{linear height increase} (for short LHI) if
there exists a number $c$ such that for every input tree
$s\in T_\Sigma$: $\he{M(s)}\leq c\cdot \he{s}$.

We now introduce two additional properties for mttrs which will allow
us to decide whether a given mttr has LSHI or LHI.
Recall that $\widehat{M}$ denotes the extension of $M$: this transducer
can translate input trees that may contain leaves that are labeled
by elements from $P$ (the set of look-ahead states of $M$).
Whenever the state $q$ of $M$, of rank $m$, encounters an input node $u$
labeled by an element $p$ of $P$, the transducer $\widehat{M}$ outputs 
$\< q,p>(y_1,\dots,y_m)$.
%, where 
%$\underline{q}\in\underline{Q}$ is a new output symbol of rank $m$
%and $t=\text{rev}(u)$ is a monadic tree that represents the reverse Dewey notation
%of the node $u$.
We call a tree in $s\in T_\Sigma(P)$ a \emph{$\Sigma$-context} if it contains
exactly one occurrence $u$ of an element of $P$.

We say that the mttr $M$ is \emph{finite nesting} (for short fnest), 
if there exists a number $c$ such that for every $\Sigma$-context $s$
there are at most $c$-many occurrences of symbols
$\<q,p>$ with $q\in Q$ on any path of the tree $\widehat{M}(s)$;
in this case, we say that $c$ is a \emph{nesting bound} of $M$. 
%We say that $M$ is \emph{infinite nesting} when it is not finite nesting. 
We say that $M$ is \emph{finite yield nesting} (for short fynest), 
if there exists
a number $c$ such that for every input tree $s\in T_\Sigma(P)$ 
there are at most $c$-many occurrences of symbols 
from $\< q,p>$ with $q\in Q$ on any path of the 
tree $\widehat{M}(s)$;
in this case, we say that $c$ is a \emph{yield nesting bound} of $M$. 
%We say that $M$ is \emph{infinite yield nesting} when it is not 
%finite yield nesting. 
The proof of the next lemma is straightforward (by reduction to Proposition~\ref{prop:finite})
and can be found in the Appendix.

\begin{lemma}\label{lm:decidable}
Let $M$ be an mttr.
Then 
(1)~it is decidable whether or not $M$ is finite nesting and
(2)~it is decidable whether or not $M$ is finite yield nesting.
\end{lemma}

Informally the next lemma is easy to understand, e.g., for Statement~(1),
if $M$ is finite nesting with bound $c$, then a single node of an input tree
can only ``contribute'' at most $c\cdot\text{mhr}$ nodes to the height of
the output tree, where mhr denotes the maximum height of the right-hand side of
any rule of the mttr. A formal proof can be found in the Appendix.

\begin{lemma}\label{lm:easy}
Let $M$ be an mttr.
(1)~If $M$ is finite nesting, then it is of linear size-to-height increase.
(2)~If $M$ is finite yield nesting, then it is of linear height increase. 
\end{lemma}

For another tree $t$ and a $\Sigma$-context $s$, $s[t]$ denotes the tree
$s[u\leftarrow t]$.

\begin{lemma}\label{lm:nest}
Let $M$ be an mttr that is depth proper.
(1)~If $M$ is not finite nesting, 
then $M$ does not have linear size-to-height increase.
(2)~If $M$ is not finite yield nesting, 
then $M$ does not have linear height increase.
\end{lemma}
\begin{proof} 
Let $M$ be given by a tuple as usual.
To prove~(1), assume that $M$ is not fnest.
We will show that this implies that $M$ does not have LSHI.
Since $M$ is not fnest (and has only finitely many states)
there must be some state $q\in Q^{(m)}$ with $m\geq 1$
that occurs arbitrarily often on paths of output trees of $\widehat{M}$.
More precisely, there are infinite sequences of contexts $c_0,c_1,\dots$
and numbers $n_0<n_1<\cdots$ such that
$q$ occurs $\geq n_0$ times on a path in $\widehat{M}(c_0)$ and
$q$ occurs $\geq n_1$ times on a path in $\widehat{M}(c_0[c_1])$, etc.
From this we can deduce (by considering sufficiently many numbers $n_i$),
similarly to the proof of Lemma~6.5 of~\cite{DBLP:journals/siamcomp/EngelfrietM03},
that $M$ is ``(nested) input pumpable'', i.e.,
there exist $q_1,q_2,j,s_0,s_1,u_0,u_1,p$ such that \vspace{-1mm}

\begin{enumerate}
\item $\< q_1,p>$ occurs in $\widehat{M}(s_0[u_0\leftarrow p])$,
\item $\widehat{M}_{q_1}(s_1[u_1\leftarrow p])$ has either: a subtree 
$\< q_1,p>(t_1,\dots,t_m)$ such that some $t_{j'}$ contains a 
subtree $\< q_2,p>(\xi_1,\dots,\xi_l)$
where $\xi_{j}$ contains $y_{j'}$ for some $j'\in[m]$, 
or a subtree $\< q_2,p>(t_1,\dots,t_l)$ such that $t_j$ contains a subtree 
$\< q_1,p>(\xi_1,$ $\dots,\xi_m)$, 
\item $\widehat{M}_{q_2}(s_1[u_1\leftarrow p])$ has a subtree 
$\< q_2,p>(t_1,\dots,t_l)$ such that $t_j$ contains $y_{j}$, and
\item $p=h(s_1/u_1)=h(s_1[u_1 \leftarrow p])$.
\end{enumerate}%\vspace{-1mm}

By ``pumping'', i.e., considering 
$s_n=s_0[u_0\leftarrow s_1[u_1\leftarrow s_1[u_1\leftarrow \dots ]]]$
with $n$ replacements of the node $u_1$, we obtain that
$\widehat{M}_{q_1}(s_n)$ contains a path with $\geq n$ nested occurrences of $\< q_2,p>$. 
Note that this proof is simpler than that of 
Lemma~6.5 of~\cite{DBLP:journals/siamcomp/EngelfrietM03} because we only look here 
at the height of outputs instead of the size of outputs. This is simpler because, 
in a mttr, a state call can copy a parameter containing large outputs of other state 
calls, creating a size growth of the output that is difficult to track, but these 
copies cannot be copied vertically on top of each other, so the output height is 
easier to track. 

Assume now by contradiction that $M$ has LSHI, i.e., there exists a $c$ such that
for every input tree $s\in T_\Sigma$: $\he{M(s)}\leq c\cdot |s|$.
Since $M$ is depth proper, we may choose $s\in L_p$ such that
$M_{q_2}(s)$ contains an occurrence of $y_{j}$ at depth
$\geq c c_1 +1$, where $c_1=|s_1[u_1\leftarrow p]|-1$.
We know that $\widehat{M}_{q_1}(s_n)$ contains $\geq n$ nested occurrences of $q_2$
(where always the $j$-th subtree of $q_2$ contains further nested occurrences of $q_2$).
Now let $t_n=s_0[u_0\leftarrow s_n[u_1^n\leftarrow s]]$
and take $n>c(c_0+c_2)$, where $c_0=|s_0[u_0\leftarrow p]|-1$ and $c_2=|s|$.
Since $|t_n|=c_0+nc_1+c_2$, we obtain that $\he{M(t_n)}>c\cdot |t_n|$ because
$\he{M(t_n)}\geq n(cc_1+1) > ncc_1 + c(c_0+c_2)=c\cdot |t_n|$ by the choice of $n$. 
So \emph{nested input pumpability} implies that $M$ is not of LSHI. 

%long version proof:
%(\emph{Detailed proof}.)
We now prove that if $M$ is not finite nesting, then it must be \emph{nested input pumpable}. 
In order to do so, we first introduce a few notations and characterize 
\emph{nested input pumpability} and the \emph{finite nesting} property using these notations. 
Let $c$ be a $\Sigma$-context and $q \in Q$ a state of $M$. 
To talk about the nesting of states in $\widehat{M}_q(c)$, we first give a notation for paths: 
\begin{enumerate}
\item For any node $u$ at depth $n$ in $\widehat{M}_q(c)$, we note the path to node $u$ as the sequence of pairs: 
\[(\ell_1,i_1) \, (\ell_2,i_2) \dots (\ell_{n},i_{n}) \, (\ell_{n+1},\bot)\] 
where $i_1, \dots, i_n$ are indexes such that $u = i_1\, i_2\dots i_n$ and, for all $j \leq n+1$, $\ell_j$ is the label of node $i_1\dots i_{j-1}$ or, if node $i_1\dots i_{j-1}$ is labeled by a state call $\< q',p>$, then $\ell_j = q'$, 
\item Since we are only interested in the nesting of states, we remove from such paths all pairs $(\ell_j,i_j)$ where $\ell \in \Delta$. We obtain nesting sequences of the form: 
\[(q_1,k_1)\, (q_2, k_2) \dots (q_n,k_n)\, (\ell_{n+1},k_{n+1})\]
where $\ell_{n+1}$ is either a state in $Q$ or a parameter, $k_{n+1} \in \{\bot\} \cup \N$, 
and for all $j \leq n$, $k_j \in [m_j]$ where $m_j$ is the arity of state $q_j$. 
\item For each such sequence, if $\ell_{n+1} = q_{n+1} \in Q$ then we write: 
\[(q,\bot) \to_c (q_1,k_1)\, (q_2, k_2) \dots (q_n,k_n)\, (\ell_{n+1},k_{n+1})\]
Otherwise $\ell_{n+1} = y_k$ is a parameter of $q$, $k_{n+1} = \bot$ and we write: 
\[(q,k) \to_c (q_1,k_1)\, (q_2, k_2) \dots (q_n,k_n)\]
\end{enumerate}
This defines a relation $\to_c \subseteq \QY \times \QY^*$ where $\QY = \{ (q,k) \mid q \in Q^{(m)}, k \in [m] \cup \{\bot\}\}$ and $\QY^*$ denotes the set of (possibly empty) sequences of elements of $\QY$. 
Note that if $(q,\bot) \to_c w$, then in the nesting sequence $w\in \QY^*$ only the last pair may contain a $\bot$. 
A \emph{nesting loop} is given by a $\Sigma$-context $c$ with a leaf labeled $p$ such that $h(c) = p$, and two pairs $(q_1,k_1), (q_2,k_2) \in \QY$ such that:
\begin{itemize}
\item $(q_1,k_1) \to_c w_1 \, (q_1,k_1) \, w_2 \, (q_2,k_2) \, w_3$ or $(q_1,k_1) \to_c w_1 \, (q_2,k_2) \, w_2 \, (q_1,k_1) \, w_3$, 
\item $(q_2,k_2) \to_c w_4 \, (q_2,k_2) \, w_5$
\item $(q_1,p)$ is reachable, i.e., there exists $\Sigma$-context $c_0$ with a leaf labeled $p$ such that $\< q_1,p>$ appears in $\widehat{M}(c_0)$. 
\end{itemize}
for some nesting sequences $w_1, w_2, w_3, w_4, w_5 \in \QY^*$. This allows us to rephrase 
the \emph{nested input pumpability} property as the existence of a \emph{nesting loop}. 
We want to prove that if $M$ is not finite nesting then it has a nesting loop. 

We extend the relation $\to_c$ to sequences of pairs on the left so that, for pairs $(q_1,k_1),\ab (q_2,k_2) \in \QY$ and sequences $w_1, w_2 \in \QY^*$, if $(q_1,k_1) \to_c w_1$ and $(q_2,k_2) \to_c w_2$ then $(q_1,k_1)\, (q_2,k_2) \to_c w_1 w_2$. More generally, for all sequences $w_1, w'_1, w_2, w'_2 \in \QY^*$, if $w_1 \to_c w'_1$ and $w_2 \to_c w'_2$ then $w_1 \, w_2 \to_c w'_1 \, w'_2$. We can now show the following claim:

\begin{claim}\label{cla:nesting_decomposition}
For all $\Sigma$-contexts $c$ and $c'$ with leafs labeled resp.\ $p$ and $p'$ such that $p=h(c')$, we can define the $\Sigma$-context $c \cdot c' = c[p \leftarrow c']$ and, for all sequences $w, w'' \in \QY^*$, if $w \to_{c\cdot c'} w''$ then there exists a sequence $w' \in \QY^*$ such that $w \to_c w' \to_{c'} w''$. 
\end{claim}

\begin{proof}
%(Sketch.) 
We only need to show this for $w = (q_0,k_0) \in \QY$ because of the definition of $\to_c$ on sequences of pairs. 
%We do so by looking at the path in $\widehat{M}_{q_1}(c \cdot c')$ reducing to $w''$. Then the corresponding path in $\widehat{M}_{q_0}(c)$ reduces to $w'$. 
%There are two cases depending on $k_0 \in [m] \cup \{\bot\}$. 
%Let us first assume that $k_0 = \bot$. 
%We note $w'' = (q_1,k_1)\, (q_2, k_2) \dots (q_n,k_n) \, (q_{n+1},\bot)$. 
%Because $(q_0,k_0) \to_{c \cdot c'} (q_1,k_1) \dots (q_n,k_n) \, (q_{n+1},\bot)$, 
Because $(q_0,k_0) \to_{c \cdot c'} w''$, there must be a path $\Pi$ in $\widehat{M}_{q_0}(c \cdot c')$ reducing 
to $w''$ (by removing pairs in $\Delta \times \N$ and removing $(y_{k_0},\bot)$ if $k_0 \neq \bot$). 
Because $\widehat{M}_{q_0}(c \cdot c') = \widehat{M}_{q_0}(c)[\< q,p> \leftarrow \widehat{M}_{q}(c')]$, 
path $\Pi$ can be similarly obtained from a path $\Pi'$ in $\widehat{M}_{q_0}(c)$ by substituting 
each $(q,k)$ with a path in $\widehat{M}_q(c')$. 
More specifically, noting $(q_1,k_1), \dots, (q_n,k_n)$ the pairs in path $\Pi'$ that are in $\QY$ 
(in order of apparition in $\Pi'$), we substitute in $\Pi'$:
\begin{itemize}
\item each occurrence of a pair $(q_i,k_i) \in \QY$ by a path $\Pi'_i$ such that $\Pi' (y_{k_0,\bot})$ is a path in $\widehat{M}_{q_i}(c')$ (for $i \leq n$), 
\item each occurrence of a pair $(q_n,\bot)$ by a path $\Pi'_n$ in $\widehat{M}_{q_n}(c')$. 
\end{itemize}
We get: $\Pi = \Pi'[(q_i,k_i) \leftarrow \Pi'_i]$ and, by removing pairs in $(\Delta \times \N) \cup (Y^m \times \{\bot\})$: 
\[w'' = w'_1\, w'_2 \, \dots \, w'_n\]
where for all $i \leq n$, $w'_i$ is obtained from $\Pi'_i$ by removing pairs in $(\Delta \times \N) \cup (Y^m \times \{\bot\})$. 
Then, for all $i \leq n$ and by definition of $\Pi'_i$, we have $(q_i,k_i) \to_{c'} w'_i$. 
So $(q_1,k_1) \dots (q_n,k_n) \to_{c'} w'_1\, w'_2 \, \dots \, w'_n = w''$. 

We note $w'$ the sequence obtained from $\Pi'$ by removing pairs in 
$(\Delta \times \N) \cup (Y^m \times \{\bot\})$.
Then $w' = (q_1,k_1) \dots (q_n,k_n)$ and so $(q_0,k_0) \to_{c} w' \to_{c'} w''$. 
%If $k_0 = \bot$ then we have $w' = (q_1,k_1) \dots (q_n,k_n)$ with $k_n = \bot$, 
%and so $(q_0,k_0) \to_{c} w' \to_{c'} w''$. 
%If $k_0 \neq \bot$ then we have $w' = (q_1,k_1) \dots (q_n,k_n)\, (y_{k_0},\bot)$, 
%and so $(q_0,k_0) \to_{c} (q_1,k_1) \dots (q_n,k_n) \to_{c'} w''$. 
\end{proof}
We could also prove that $\to_{c\cdot c'} \,=\, \to_{c'} \circ \to_{c}$, but it is not necessary for this proof. 

To prove that $M$ has a nesting loop (i.e.\ $M$ is nested input pumpable), we assume that $M$ is not finite nesting. 
Then, for all $n \in \N$, there exists a $\Sigma$-context $c_n$ such that: $(q_0,\bot) \to_{c_n} w$ for some $w \in \QY^*$ with $|w| \geq n$. We can decompose any such $c_n$ into a concatenation $c_{n,1} \cdot c_{n,2} \cdot \dots c_{n,r}$ and use the claim to obtain: 
\[(q_0,\bot) \to_{c_{n,1}} w_1 \to_{c_{n,2}} w_2 \dots \to_{c_{n,r}} w_r \]
where $w_1, w_2, \dots w_r \in \QY^*$ and $|w_r| = |w| \geq n$. By choosing a big enough $n$, we will show how to find a \emph{nesting loop}. To do that, we decompose $c_n$ into several contexts and use the claim. 

A $\Sigma$-context $c$ is \emph{atomic} if its leaf labeled in $P$ is a child node of its root. 
Let $c = \sigma(t_1, \dots, t_{i-1}, p_i, t_{i+1}, \dots, t_k)$ be an atomic $\Sigma$-context and $q \in Q$ a state of $M$. Noting $p_j = h(t_j)$ for $j \neq i$, there is in $M$ a rule $\< q, \sigma(x_1:p_1, \dots, x_k:p_k)> (y_1, \dots, y_m) \to t$. 
Then $\widehat{M}_q(c) = t[\< q', x_j> \leftarrow M_{q'}(c/j) \mid j \neq i]$. Because $M_{q'}(c/j) \in T_\Delta$ for all $q' \in Q$ and $j \neq i$, the nesting of state calls in $\widehat{M}_q(c)$ is the nesting of state calls of the form $\< q',x_i>$ in $t$. 
So, for $(q,k) \in \QY$, the length of nesting sequences $w$ such that $(q,k) \to_c w$ is bounded by the height of $t$. 
There is a finite number of rules for $M$, so there is a finite number of such $t$ and the length of sequences $w \in \QY$ such that $(q,k) \to_c w$ has an upper bound $B$ that does not depend on $q,k$ or $c$. 
In other words, for all $(q,k) \in \QY$, $w \in \QY^*$ and atomic $\Sigma$-context $c$ we have:
\[ (q,k) \to_c w ~~~~ \Rightarrow ~~~~ |w| \leq B \]
Moreover, for all $w_1, w_2 \in \QY^*$: $w_1 \to_c w_2 ~~~ \Rightarrow ~~~ |w_2| \leq B\, |w_1|$. 

We decompose the $\Sigma$-context $c_n$ into atomic $\Sigma$-contexts $c_{n,1}, c_{n,2}, \dots, c_{n,r}$. 
Since $(q_0,\bot)\ab \to_{c_{n,1}} w_1 \to_{c_{n,2}} w_2 \, \dots \to_{c_{n,r}} w_r$, 
we have $|w_r| \leq B^r$ and, since $|w_r| \geq n$: $n \leq B^r$. 
So, by taking $n$ big enough, we can also make $r$ as big as we want. 
In order to find a nesting loop, we require more structure on the nesting sequences 
$c_{n,1}, \dots, c_{n,r}$. The precise structure we need is described in the following claim: 
\begin{claim}
For all for all $r \in \N$, if there exists a $\Sigma$-context $c$, a pair $\theta \in \QY$ and a sequence $w \in \QY^*$ with $\theta \to_{c} w$ and $|w| \geq B^r$, then there exists $\Sigma$-contexts $c_1, \dots, c_{r-1}$, look-ahead states $p_1, \dots, p_r$ and pairs $\theta_{i,j} \in \QY$ for all $i,j \in [r]$ with $j \leq i$ such that:
\begin{itemize}
\item for all $i \in [r-1]$, $h(c_i) = p_i$ and $c_i$ has a leaf labeled $p_{i+1}$, 
\item $\theta_{1,1} = \theta$, 
\item for all $i,j \in [r-1]$ with $j < i$, there exists $w_{i,j}, w'_{i,j} \in \QY^*$ such that: 
$\theta_{i,j}\ab \to_{c_i} w_{i,j} \, \theta_{i+1,j} \, w'_{i,j}$, 
\item for all $i \in [r-1]$, there exists $w_i, w'_i, w''_i \in \QY^*$ such that either 
$\theta_{i,i} \to_{c_i} w_{i} \, \theta_{i+1,i}\ab \, w'_{i} \, \theta_{i+1,i+1} \, w''_i$ or 
$\theta_{i,i} \to_{c_i} w_{i} \, \theta_{i+1,i+1} \, w'_{i} \, \theta_{i+1,i} \, w''_i$. 
\end{itemize}
\end{claim}

These conditions can be summed up graphically. To simplify the picture, we replace all sequences 
$w_{i,j}, w'_{i,j}, w_i, w'_i, w''_i$ for $i,j \in [r]$ with the symbol $\thicksim$. 
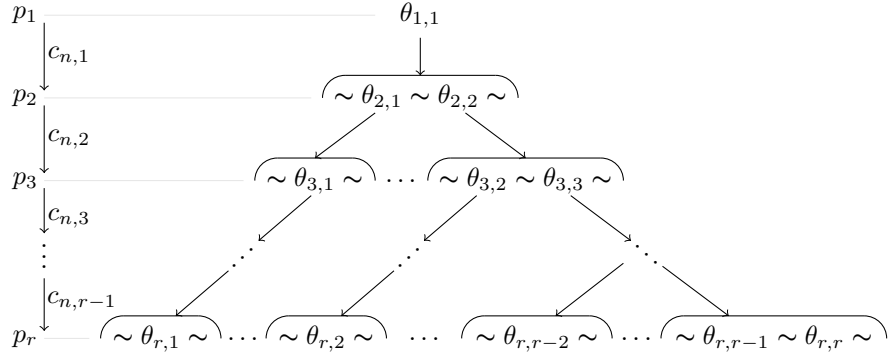
\begin{center}
	\begin{tikzpicture}
		\newcommand{\halfblob}[5]{ % #1: radius, #2: half-length, #3: x-coordinate, #4: y-coordinate, #5: name
			\draw (#3 - #2,#4 + #1) arc [start angle=90, end angle=180, radius=#1];
			\draw (#3 + #2 + #1,#4) arc [start angle=0, end angle=90, radius=#1];
			\draw (#3 - #2,#4 + #1) -- (#3 + #2,#4 + #1);
			%\draw (#3 - #2,#4 - #1) -- (#3 + #2,#4 - #1);
			\node at (#3,#4) {#5};
		}
		
		\node at (0,0) {$\theta_{1,1}$};
		\draw[->] (0,-0.3) -> (0,-0.8);
		\halfblob{0.3}{1}{0}{-1.1}{$\thicksim \theta_{2,1} \thicksim \theta_{2,2} \thicksim$}
		
		\draw[->] (-0.6,-1.3) -> (-1.4,-1.9);
		\halfblob{0.3}{0.5}{-1.4}{-2.2}{$\thicksim \theta_{3,1} \thicksim$}
		\draw[->] (0.6,-1.3) -> (1.4,-1.9);
		\halfblob{0.3}{1}{1.4}{-2.2}{$\thicksim \theta_{3,2} \thicksim \theta_{3,3} \thicksim$}
		\node at (-0.23,-2.2) {$\dots$};
		
		\draw[->] (-1.45,-2.4) -> (-2.15,-3);
		\node[rotate=43] at (-2.34,-3.2) {$\dots$};
		\draw[->] (-2.55,-3.4) -> (-3.25,-4);
		\halfblob{0.3}{0.5}{-3.45}{-4.3}{$\thicksim \theta_{r,1} \thicksim$}
		
		\draw[->] (0.75,-2.4) -> (0.05,-3);
		\node[rotate=43] at (-0.13,-3.2) {$\dots$};
		\draw[->] (-0.35,-3.4) -> (-1.05,-4);
		\halfblob{0.3}{0.5}{-1.25}{-4.3}{$\thicksim \theta_{r,2} \thicksim$}
		
		\draw[->] (2,-2.4) -> (2.8,-3);
		\node[rotate=-40] at (3.03,-3.2) {$\dots$};
		\draw[->] (2.7,-3.3) -> (1.8,-4);
		\halfblob{0.3}{0.7}{1.55}{-4.3}{$\thicksim \theta_{r,r-2} \thicksim$}
		\draw[->] (3.22,-3.34) -> (4.1,-4);
		\halfblob{0.3}{1.2}{4.7}{-4.3}{$\thicksim \theta_{r,r-1} \thicksim \theta_{r,r} \thicksim$}
		
		\node at (-2.33,-4.3) {$\dots$};
		\node at (0.08,-4.3) {$\dots$};
		\node at (2.9,-4.3) {$\dots$};
		
		%vertical arrows on the side
		\draw[->] (-5,-0.1) -> (-5,-1);
		\node at (-4.67,-0.55) {$c_{n,1}$};
		\draw[->] (-5,-1.2) -> (-5,-2.1);
		\node at (-4.67,-1.65) {$c_{n,2}$};
		\draw[->] (-5,-2.3) -> (-5,-2.9);
		\node at (-4.67,-2.7) {$c_{n,3}$};
		\node at (-5,-3.1) {$\vdots$};
		\draw[->] (-5,-3.5) -> (-5,-4.2);
		\node at (-4.5,-3.8) {$c_{n,r-1}$};
		
		%horizontal gray lines
		\draw[gray!20] (-5,0) -> (-0.6,0);
		\draw[gray!20] (-5,-1.1) -> (-1.45,-1.1);
		\draw[gray!20] (-5,-2.2) -> (-2.35,-2.2);
		\draw[gray!20] (-5,-4.3) -> (-4.4,-4.3);
		
		%look-ahead states
		\node at (-5.25,0) {$p_1$};
		\node at (-5.25,-1.1) {$p_2$};
		\node at (-5.25,-2.2) {$p_3$};
		\node at (-5.25,-4.3) {$p_r$};		
		
	\end{tikzpicture}
\end{center}
Note that, in this representation, we chose to represent 
$\theta_{i,i} \to_{c_i} w_{i} \, \theta_{i+1,i} \, w'_{i} \, \theta_{i+1,i+1} \, w''_i$ instead of 
$\theta_{i,i} \to_{c_i} w_{i} \, \theta_{i+1,i+1} \, w'_{i} \, \theta_{i+1,i} \, w''_i$ for all $i \in [r-1]$. 
But this distinction does not matter to the proof of the claim. 
%but because the order of nesting does not matter for nesting loops, 
%we can assume that branching is happening on the right without loss of generality (in fact 
%branching on the left makes the proof easier by excluding the case where $\bot$ occurs on the branching side). 
From now on we use $\thicksim$ to denote arbitrary sequences in $\QY^*$ which we will not use to find a nesting loop. 

\begin{proof}
We prove this by induction on $r$. 
Let $c$ be a $\Sigma$-context, $\theta$ a pair in $\QY$ and $w$ a sequence in $\QY^*$ with $\theta \to_{c} w$ and $|w| \geq B^{r+1}$. 
We split $c$ into atomic $\Sigma$-contexts $c'_1, \dots, c'_n$, then we have sequences $w_1, \dots, w_{n-1} \in \QY^*$ 
such that $\theta \to_{c'_1} w_1 \dots \to_{c'_{n-1}} w_{n-1} \to_{c'_n} w$. 
Let $i$ be the largest $i$ such that there is a pair $\theta'$ in sequence $w_i$ with $\theta' \to_{c'_{i+1} \cdots c'_n} w'$ and $|w'| \geq B^r$. 
If we had $|w'| \geq B^{r+1}$ then, because $c'_{i+1}$ is atomic, we would have a $\theta''$ in sequence $c'_{i+1}$ with $\theta'' \to_{c'_{i+2} \cdots c'_n} w''$ and $|w''| \geq B^r$. 
So $B^r \leq |w'| < B^{r+1} \leq |w|$. 
Therefore there is another pair $\theta_{2,1}$ in $w_i$ (other than $\theta'$) with $\theta_{2,1} \to_{c'_{i+1} \cdots c'_n} w''$ and $|w''| \geq 1$. 

Since $\theta' \to_{c'_{i+1} \cdots c'_n} w'$ and $|w'| \geq B^r$, we use the induction hypothesis on $theta'$ and $c'_{i+1} \cdots c'_n$. In order to prove the induction for $r+1$, we rename the $\Sigma$-contexts $c_1, \dots c_{r-1}$, look-ahead states $p_1, \dots, p_r$ and pairs $\theta_{i,j}$ (for $j \leq i \leq r$) into $\Sigma$-contexts $c_2, \dots c_{r}$, look-ahead states $p_2, \dots, p_{r+1}$ and pairs $\theta_{i+1,j+1}$ (for $j \leq i \leq r$). 
Then $c'_{i+1} \cdots c'_n = c_2 \cdots c_{r}$. 

Since $\theta_{2,1} \to_{c_2 \cdots c_r} w''$ with $|w''| \geq 1$, there are pairs $\theta_{3,1}, \dots, \theta_{n,1}$ such that $\theta_{n,1}$ appears in sequence $w''$ and, for all $i \in [r]$ with $i \geq 2$, $\theta_{i,1} \to_{c_i} w'_i \, \theta_{i+1,1} \, w''_i$ with $w'_i, w''_i \in \QY^*$. 
To conclude, we choose $c_1 = c'_1 \dots c'_i$, $p_1 = h(c_1)$ and $\theta_{1,1} = \theta$. 
\end{proof}

In order to find a nesting loop, we need two indexes $i$ and $j$ with: 
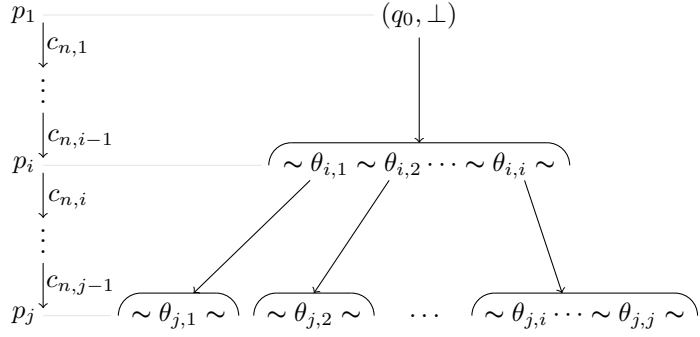
\begin{center}
	\begin{tikzpicture}
		\newcommand{\halfblob}[5]{ % #1: radius, #2: half-length, #3: x-coordinate, #4: y-coordinate, #5: name
			\draw (#3 - #2,#4 + #1) arc [start angle=90, end angle=180, radius=#1];
			\draw (#3 + #2 + #1,#4) arc [start angle=0, end angle=90, radius=#1];
			\draw (#3 - #2,#4 + #1) -- (#3 + #2,#4 + #1);
			%\draw (#3 - #2,#4 - #1) -- (#3 + #2,#4 - #1);
			\node at (#3,#4) {#5};
		}
		
		\node at (0,0) {$(q_0,\bot)$};
		\draw[->] (0,-0.3) -> (0,-1.7);
		\halfblob{0.3}{1.7}{0}{-2}{$\thicksim \theta_{i,1} \thicksim \theta_{i,2} \dots \thicksim \theta_{i,i} \thicksim$}
		
		\draw[->] (-1.45,-2.2) -> (-3,-3.7);
		\halfblob{0.3}{0.5}{-3.2}{-4}{$\thicksim \theta_{j,1} \thicksim$}
		
		\draw[->] (-0.4,-2.2) -> (-1.4,-3.7);
		\halfblob{0.3}{0.5}{-1.4}{-4}{$\thicksim \theta_{j,2} \thicksim$}
		
		\draw[->] (1.4,-2.2) -> (1.9,-3.7);
		\halfblob{0.3}{1.2}{2.2}{-4}{$\thicksim \theta_{j,i} \dots \thicksim \theta_{j,j} \thicksim$}
		\node at (0.1,-4) {$\dots$};
		
		%vertical arrows on the side
		\draw[->] (-5,-0.1) -> (-5,-0.7);
		\node at (-4.67,-0.43) {$c_{n,1}$};
		\node at (-5,-0.9) {$\vdots$};
		\draw[->] (-5,-1.3) -> (-5,-1.9);
		\node at (-4.5,-1.65) {$c_{n,i-1}$};

		\draw[->] (-5,-2.1) -> (-5,-2.7);
		\node at (-4.67,-2.45) {$c_{n,i}$};
		\node at (-5,-2.9) {$\vdots$};
		\draw[->] (-5,-3.3) -> (-5,-3.9);
		\node at (-4.5,-3.6) {$c_{n,j-1}$};
		
		%horizontal gray lines
		\draw[gray!20] (-5,0) -> (-0.6,0);
		\draw[gray!20] (-5,-2) -> (-2.1,-2);
		\draw[gray!20] (-5,-4) -> (-4.1,-4);
		
		%look-ahead states
		\node at (-5.25,0) {$p_1$};
		\node at (-5.25,-2) {$p_i$};
		\node at (-5.25,-4) {$p_j$};		
		
	\end{tikzpicture}
\end{center}
Formally, we require two indexes $i,j$ with $i < j < r$ which share the same:
\begin{itemize}
\item look-ahead state $h(c_{n,i}) = h(c_{n,j})$, 
\item pair $\theta_{i,i} = \theta_{j,j} \in \QY$, 
\item set of pairs $\{\theta_{i,\ell}\}_{0\leq \ell \leq i} = \{\theta_{j,\ell}\}_{0\leq \ell \leq j}$.  
\end{itemize}
We ensure the existence of such $i,j$ by taking $r \geq |P|\,|Q|\,(m+1) \, 2^{|Q| (m+1)} +1$ 
%(so $n = B^{|P|\,|Q|\,(m+1)\, 2^{|Q| (m+1)} +1}$) 
where $m$ is the maximum arity of states. 
%We ensure the existence of such $i,j$ by taking $r \geq |P|\,|\QY| \, 2^{|\QY|} +1$ 
%(so $n = B^{|P|\,|\QY|\, 2^{|\QY|} +1}$) with $|\QY| \leq |Q| \, (m+1)$ where $m$ is the maximum arity of states. 
We now show how to build the nesting loop from indexes $i,j$. 
We note $p = h(c_{n,i}) = h(c_{n,j})$, $(q_1,k_1) = \theta_{i,i} = \theta_{j,j}$ and 
$S = \{\theta_{i,\ell}\}_{0\leq \ell \leq i} = \{\theta_{j,\ell}\}_{0\leq \ell \leq j}$. 
We note $c' = c_{n,i}.c_{n,i+1}. \dots. c_{n,j-1}$. 
Note that $c'$ has a leaf labeled $p$ and $h(c')=p$. 
%drawing maybe

We need the sets $\{\theta_{i,\ell}\}_{0\leq \ell \leq i}$ and $\{\theta_{j,\ell}\}_{0\leq \ell \leq j}$ 
to be equal so that the pairs $\theta_{i,k}$ for $k \leq i$ loop on each other through the loop $c'$. 
Formally, noting $\theta'_0 = \theta_{j,i}$, for all $\theta'_k \in S$ for $k \in \N$, there exists 
$\theta'_{k+1} \in S$ such that $\theta'_k \to_c' \,\thicksim \theta'_{k+1} \thicksim$. 
Since $S \subseteq \QY$ is finite, there must be $n, m \in \N$ such that 
$\theta'_{n} = \theta'_{n+m}$ (with $m \geq 1$), so $\theta'_n \to_{c'^m} \,\thicksim \theta'_n \thicksim$. 
Also $(q_1,k_1) \to_{c'} \,\thicksim \theta'_0 \thicksim (q_1,k_1) \thicksim\,$ and $\theta'_0 \to_{c'^n} x_n$, 
so $(q_1,k_1) \to_{c'^{n+1}} \, \thicksim \theta'_n \thicksim (q_1,k_1) \thicksim\,$ and, for all $m' \in \N$: 
$(q_1,k_1) \to_{c'^{m'}} \, \thicksim (q_1,k_1) \thicksim \, \to_{c'^{n+1}} \, \thicksim \theta'_n \thicksim (q_1,k_1) \thicksim$. 
Finally, for $c = c'^{m(n+1)}$, we have 
$(q_1,k_1) \to_{c} \, \thicksim \theta'_n \thicksim (q_1,k_1)\thicksim$ and $\theta'_n \to_{c} \theta'_n$. 
So we have a \emph{nesting loop}. 

In conclusion, if $M$ is not finite nesting, then it is \emph{nested input pumpable}, and so it does not have linear size-to-height increase. 
\vspace{2mm}

The proof of Statement~(2) can be given in a very similar way as for~(1), here we only outline the changes to the notations which allow to adapt the proof of~(1) to~(2). 
We replace $\Sigma$-contexts with elements of the set $T_\Sigma(P)$ containing possibly several leafs labeled in $P$. The rest of the notational changes are consequences of this change. 
Given a $s \in T_\Sigma(P)$, we now consider the nesting of state calls called on distinct subtrees of $s$ with potentially distinct look-ahead states. We augment pairs in $\QY$ so as to include the look-ahead, so $\QY = \{ (q,k,p) \mid q \in Q^{(m)}, k \in [m] \cup \{\bot\}, p \in P\}$. The notation $(q,k,p) \to_s (q_1,k_1,p_1) \dots (q_n,k_n,p_n)$ means that $h(s)=p$ and calls to states $q_1, \dots, q_n$ on leafs of $s$ labeled $p_1, \dots, p_n$ resp.\ are nested on parameters $y_{k_1}, \dots, y_{k_n}$ along a path in $\widehat{M}_q(s)$. This means that, when concatenating contexts, we write $s(s_1, \dots, s_m)$ instead of $s\cdot s'$. 

For~(2), similarly to nesting loops for~(1), we define a \emph{yield nesting loop} as given by contexts $s_1, s_2 \in T_\Sigma(P)$, look-ahead states $p_1, p_2 \in P$ and triplets $(q_1,k_1,p_1), (q_2,k_2,p_2) \in \QY$ such that:
\begin{itemize}
\item $h(s_1) = p_1$, $h(s_2) = p_2$, $s_1$ has two leafs labeled $p_1$ and $p_2$, $s_2$ has one leaf labeled $p_2$, 
\item $\<q_1,p_1>$ is reachable, 
\item either $(q_1,k_1,p_1) \to_{s_1} \thicksim\, (q_2,k_2,p_2) \,\thicksim\, (q_1,k_1,p_1) \,\thicksim$ \\
\phantom{.} \hspace{2.5mm} or $(q_1,k_1,p_1) \to_{s_1} \thicksim\, (q_1,k_1,p_1) \,\thicksim\, (q_2,k_2,p_2) \,\thicksim$,
\item $(q_2,k_2,p_2) \to_{s_2} \thicksim\, (q_2,k_2,p_2) \,\thicksim$. 
\end{itemize}
  
We say that $M$ is \emph{yield nested input pumpable} when it has either a \emph{yield nesting loop} or a \emph{nesting loop}. 
Note that the existence of either of these loops falsifies the linear height increase property. 
To prove~(2) we prove that infinite yield nesting implies the existence of either a yield nesting loop or a nesting loop. That proof works similarly to~(1): $M$ is not fynest so we can find large enough nesting sequences (but with the new definition of $\to_s$), find a repeating triplet $(q_1,k_1,p_1)$, pump the loop enough times that a triplet $(q_2,k_2,p_2)$ loops onto itself. Note that if the nested calls to $(q_1,k_1,p_1)$ and $(q_2,k_2,p_2)$ in $(q_1,k_1,p_1) \to_{s_1} \,\thicksim (q_2,k_2,p_2) \thicksim (q_1,k_1,p_1) \thicksim$ are on the same leaf in $s_1$ (with $p_1 = p_2$), then we get a nesting loop (otherwise we get a yield nesting loop).

\end{proof}

From Theorem~\ref{th:proper} and Lemmas~\ref{lm:decidable},~\ref{lm:easy}, and~\ref{lm:nest} we obtain
our following main theorem.

\begin{theorem}
Let $M$ be an mttr. 
Then 
(1)~it is decidable whether or not $M$ has linear size-to-height increase
(2)~it is decidable whether or not $M$ is linear height-increase.
\end{theorem}

\section{Conclusions}

We have proven that for a given macro tree transducer (with look-ahead) it
is decidable whether or not it has linear height increase (LHI) and,
whether or not it has linear size-to-height increase (LSHI).
Both decision procedures rely on a novel normal form that is called
``depth-proper'' normal form. Roughly speaking the normal form requires
that each parameter of every state of the transducer appears at
arbitrary large depths in output trees generated by that state (and for a given
look-ahead state). 

One major open problem in the field is to prove a Conjecture of Joost Engelfriet
(from around the year 2000), that the translation of an mttr can be defined by
an attributed tree transducer (atts) if and only if the translation 
has ``linear size to number of distinct output subtrees'' increase.
Note that deciding such property is out of reach (it is at least as difficult as deciding
equivalence of atts). To prove this characterization, different loops need to
be considered which produce unbounded numbers of states in (partial) output trees.
We believe that the depth normal form will be instrumental in reducing the number of
different such loops that must be considered and therefore will be of great help
in proving the conjecture.

\bigskip

%\subsection*{Acknowledgements}
\textbf{Acknowledgements.}\quad
We thank the anonymous reviewers of a previous version of this paper
for their very careful and critical comments which hugely helped to improve the paper.

\newpage

%CASE: COMPOSITION OF TOTAL TRANSDUCERS\\
%better time complexity as modification of $T_1$ and look-ahead are not necessary.\\
%Construction of  $M$, $O(T_1^{T_2})$

\bibliographystyle{splncs04}% the mandatory bibstyle
\bibliography{mybib}

\newpage
\appendix
%!TEX root = ICALP_main.tex
\section*{Appendix}

\subsection*{Proof of Lemma~\ref{lm:nd}}

Statement of the lemma:
Let $\Delta$ be a ranked alphabet, $m\geq 1$, $Z\subseteq Y_m$, and
$t\in T_\Delta(Y_m)$.
(1)~$t=\lcop{t}_Z[u\leftarrow t/u\mid u\in\mathcal{U}(\lcop{t}_Z)]$.
(2)~$\paras{\lcop{t}_Z} = \paras{t}$. 

\smallskip

\begin{proof}
The proof is by induction on the structure of $\lcop{t}_Z$.
Let us denote the substitution $[u\leftarrow t/u\mid u\in\mathcal{U}(\lcop{t}_Z)]$
by $[t]$.
There are three cases.

Case 1: $t = y\in Z$. Then $\mathcal{U}(\lcop{t}_Z)=\emptyset$.
Hence $\lcop{t}_Z[t]=\lcop{y}_Z$. The latter equals $y=t$ by 
the definition of $\lcop{.}_Z$. Thus~(1)~holds. Also 
$\paras{\lcop{t}_Z}=\{y\}=\paras{t}$ and so~(2)~holds. 
%the definition of $\lcop{.}_Z$. Thus~(1)~holds and~(2)~holds for $u=\epsilon$.

Case 2: $\paras{t} \cap Z = \emptyset$. Then $\lcop{t}_Z = \paras{t} = Z' \subseteq Y_m$ 
by the definition of $\lcop{.}_Z$, and $\mathcal{U}(\lcop{t}_Z)=\{ \epsilon \}$. Hence
$\lcop{t}_Z[t] = \lcop{t}_Z[\epsilon\leftarrow t/\epsilon = t] = t$ which proves~(1). 
Again~(2)~holds 
because $\paras{\lcop{t}_Z}=\paras{\paras{t}}=\paras{t}$. 
%for $u=\varepsilon$ for each $y \in \paras{t}$ i.e.\ $y$ occurring in $t$.

Case 3: \(\paras{t}\cap Z\ne\emptyset\). Then $t=\delta(t_1,\dots,t_n)$, 
$\delta\in\Delta^{(n)}$, $n\geq 0$,
and $t_1,\dots, t_n\in T_\Delta(Y_m)$.
To show~(1) we obtain from the definition of $\lcop{t}_Z$ that
\[
\lcop{t}_Z[t]=\delta(\lcop{t_1}_Z,\dots,\lcop{t_n}_Z)[t]
= \delta(\lcop{t_1}_Z[t_1],\dots,\lcop{t_n}_Z[t_n]),
\]
where for $i\in[n]$, $[t_i]$ denotes the substitution
$[u\leftarrow t_i/u\mid u\in\mathcal{U}(\lcop{t_i}_Z)]$.
By induction the latter equals $\delta(t_1,\dots,t_n)=t$.
Finally~(2)~is implied by the induction hypothesis:
$\paras{\lcop{t}_Z}=\bigcup_{j\in[n]} \paras{\lcop{t_j}_Z}
=\bigcup_{j\in[n]} \paras{t_j}= \paras{t}$
%To show~(2), assume $y\in Y_m$ occurs in $t$.
%Thus, $n\geq 1$ and there exists an $i\in[n]$
%such that $y$ occurs in $t_i$.
%By induction of Statement~(2), there exists a leaf $u$ of $t_i$ such that 
%Statement~(2) holds (for $u$ and $t_i$); therefore Statement~(2) holds
%for $iu$ and $t$.
%\qed
\end{proof}

\subsection*{Proof of Lemma~\ref{lm:lcop_finite}}

Statement of the lemma:
Let $M$ be an mtt, $q$ a state of $M$, and $p$ a look-ahead state of $M$
such that $Y(q,p)\not=\emptyset$.
Then $\lcop{M_q(L_p)}_{Y(q,p)}$ is finite.

\smallskip

\begin{proof}
Assume that $Y_{q,p}=\{y_{j_1},\dots,y_{j_n}\}$ where $n\geq 1$.
It follows from the definition of $Y(q,p)$ that for each $i\in[n]$ there exists
a number $d_i$ such that $y_{j_i}$ occurs at depth $\leq d_i$ in 
any tree in $M_q(L_p)$. Let $d$ be the maximum of all numbers in
$\{ d_1,\dots,d_n\}$. Then every parameter in $Y(q,p)$ occurs at depth $\leq d$ in
any tree in $M_q(L_p)$. It follows from the definition of $U=\lcop{M_q(L_p)}_{Y(q,p)}$
that every node of a tree in $U$ has depth $\leq d$. Thus, $U$ is finite.
%\qed
\end{proof}

\subsection*{Example of Constructing a Depth Proper MTTR}

Let \(M=(\{q_0,q_1,q_2,q_3,q_4,\qid\},\{p\},\Sigma,$ $\Delta,q_0,R,h_0)\)
with \(\Sigma=\{a^{(1)},e^{(0)}\}\) and 
%\(\Delta=\{a_1^{(1)},a_2^{(2)},a_3^{(2)},a_4^{(1)},f^{(2)},g^{(3)},e^{(0)}\}\)
\(\Delta=\{a_1^{(1)},a_2^{(1)},\ab f_1^{(2)},f_2^{(2)},f_3^{(2)},g^{(3)},e^{(0)}\}\)
be an mttr 
where \((\Sigma,\{p\},h_0)\) is the trivial look-ahead with
$L_p=T_\Sigma$,
Suppose that \(R\) contains the following rules for \(a\): 
\begin{align*}
\<q_0, a(x)> &\to \<q_1,x>(e)
\\
\<q_1, a(x)>(y) &\to f_1(y, \<q_2,x>(a_1(y)))
\\
\<q_2, a(x)>(y) &\to g(y, \<q_3,x>(f_2(y, \<\qid, x>)), \<\qid, x>)
\\
\<q_3, a(x)>(y) &\to f_1(y, \<q_4,x>(f_3(y, \<\qid, x>)))
\\
\<q_4, a(x)>(y) &\to a_2(y)
\end{align*}
Then
because of \(L_p=\tree\Sigma\)
we have \(q_i\in F_p\) and \(Y(q_i,p)=\{y\}\) for any \(i=1,2,3,4\), and
the corresponding sets of \(\{y\}\)-skeleta are
\begin{align*}
\lcop{M_{q_1}(L_p)}_{\{y\}} &= 
\{f_1(y, g(a_1(y), f_1(f_2(a_1(y), \emp), a_2(f_3(f_2(a_1(y), \emp), \emp))), \emp))\}
\\
\lcop{M_{q_2}(L_p)}_{\{y\}} &= 
\{g(y, f_1(f_2(y, \emp), a_2(f_3(f_2(y, \emp), \emp))), \emp)\}
\\
\lcop{M_{q_3}(L_p)}_{\{y\}} &= \{f_1(y, a_2(f_3(y, \emp)))\}
\\
\lcop{M_{q_4}(L_p)}_{\{y\}} &= \{a_2(y)\}
\end{align*}
when the rules for \(e\) are appropriately given
so as for each of the sets above to be a singleton for simplicity.
Let \(t_i\) be the skeleta such that \(\lcop{M_{q_i}(L_p)}_{\{y\}} = \{t_i\}\).
When the skeleton contains no sequence node such as \(t_4\),
the \(q_4\)-calls in the right-hand side of every rule are easily removed
by substituting (in a second-order fashion) \(t_4\) for \(\<q_4,x>\).
For example,  the \(q_3\)-rule above will be 
\[ 
\<q_3, a(x)>(y) \to f_1(y, a_2(f_3(y, \<\qid, x>)))\text. 
\]
When the skeleton contains sequence nodes such as $\emptyset$,
our construction replaces them by \emph{helper states}; such states
traverse the input until further output needs to be generated.
A helper state has the form \([q,p,t,u]\) where
\(\<[q,p,t,u],s>\) for \(s\in L_p\) computes 
the subtree at node \(u\) of $M_q(s)$.
For example, the \(q_2\)-rule above which contains a \(q_3\)-call will be
\begin{align*}
&\<q_2, a(x)>(y) \to\\&\qquad
     g(y, 
       f_1(f_2(y, \<\qid, x>), a_2(f_3(f_2(y,\<\qid, x>), 
                                       \<[q_3,p,t_3,212],x>))),
       \<\qid, x>)
\end{align*}
by replacing the call by \(t_3\) whose sequence node is replaced by a helper state.
Since the sequence node \(t_3/212\) is $\emptyset$, the helper state has no parameter.
We will see later the case of skeleta containing non-empty sequence nodes.
Similarly the \(q_0\)-rule will be 
\begin{align*}
&\<q_0, a(x)>\to\\&\qquad %\<q_1,x>(e)
f_1(e, g(\begin{array}[t]{l}
         a_1(e), \\
         f_1(\begin{array}[t]{l}
             f_2(a_1(e), \<[q_1,p,t_1,2212],x>), \\
             a_2(f_3(f_2(a_1(e), \<[q_1,p,t_1,222112],x>), 
                     \<[q_1,p,t_1,22212,x>))), 
             \end{array} \\
         \<[q_1,p,t_1,23],x>))
         \end{array}
\end{align*}
A rule of a helper state \([q,p,t,u]\) is constructed from the \(q\)-rule
by first replacing each call of an \emph{improper} state
(at least one of whose parameter is improper) in the right-hand side
by the corresponding skeleton
and then taking the subtree at \(u\).
Thereby, from the obtained \(q_1\)-, \(q_2\)-, and \(q_3\)-rules above,
our construction gives
\begin{align*}
\<[q_1,p,t_1,2212],a(x)>   &\to \<[q_2,p,t_2,212],x>   \\
\<[q_1,p,t_1,222112],a(x)> &\to \<[q_2,p,t_2,22112],x> \\
\<[q_1,p,t_1,22212],a(x)>  &\to \<[q_2,p,t_2,2212],x>  \\
\<[q_1,p,t_1,23],a(x)>     &\to \<[q_2,p,t_2,3],x>     \\
\<[q_2,p,t_2,212],a(x)>    &\to \<\qid,x>              \\
\<[q_2,p,t_2,22112],a(x)>  &\to \<\qid,x>              \\
\<[q_2,p,t_2,2212],a(x)>   &\to \<[q_3,p,t_3,212],x>   \\
\<[q_2,p,t_2,3],a(x)>      &\to \<\qid,x>              \\
\<[q_3,p,t_3,212],a(x)>    &\to \<\qid,x> \text.
\end{align*}
For example, 
the right-hand side of the first rule of \(\<[q_1,p,t_1,2212],a(x)>\)
is constructed from the right-hand side of the rule of \(\<q_1,a(x)>\)
after substituted as shown above
by extracting its subtree at \(2212\), that is, \(\<[q_2,p,t_2,212],x>\).

Consider an input tree $s=a^5(s')$ with $s'\in T_\Sigma$
consisting of at least five top-most $a$-symbols
and apply the new $q_0$-rule shown above.
All the helper states in the right-hand side of that rule are now traversing
the node $1$ of $s$.
Now consider the derivation of the helper state $[q_1,p,t_1,22212]$:
it becomes $[q_2,p,t_2,2212]$ on node $11$ of $s$.
Then it become $[q_3,p,t_3,212]$ on node $111$ of $s$.
And finally it becomes $q_{\text{id}}$ on node $1111$ of $s$.
This is exactly right and corresponds to the computation of the original transducer.
All other helper states correctly become $q_{\text{id}}$ on node $111$ of $s$.

\subsection*{Proof of Lemma~\ref{lm:corr}}

Before we repeat the statement of Lemma~\ref{lm:corr} and present
its proof we need a small lemma showing that
the skeleton of the output of an mttr \(M\) can be directly computed
from given a input tree 
by modifying the rules of \(M\).
%given a 
%right-hand-side of rule $t$, the skeleton of $t$ can be computed from the 
%skeleta of state calls appearing in $t$. 
For this lemma we first need to define how to compute second-order 
substitutions of skeleta,
which will be used for the modification of the right-hand sides of rules.
We do so on a \emph{nondeleting} mttr $M$, 
i.e.\ such that states always use all their parameters. 

\begin{definition}\label{def:meta-skeleta}
%Let $M$ be a nondeleting mttr as before. 
  %
Let $\Gamma$ be a ranked alphabet
 and
let $t_1,\dots,t_n\in T_\Gamma(Y)$. 
Let $s\in T_\Gamma(Y_n\cup\mathcal{P}(Y_n))$. 
%
%\begin{itemize}
%\item
%\item Let \(k_i = 0\) for all \(i\in[n]\).
The \emph{special first-order substitution} 
$[y_i \leftarrow t_i \mid i \in [n]]\su$ (for short $[.]\su$) applied to $s$ is
inductively defined as:
\begin{align*}
	s[.]\su &= 
	\begin{cases}
		t_i &
%		\text{if \(s=\gamma_i\) for \(i\in [n]\)}
		\text{if \(s=y_i\) for \(i\in [n]\)}
		\\
		\gamma(s_1[.]\su,\dots,s_k[.]\su) &
%		\text{if \(s = \gamma(s_1,\dots,s_k)\) and }\gamma\not\in\{\gamma_1,\dots,\gamma_n\}
		\text{if \(s = \gamma(s_1,\dots,s_k)\)}
		\\
		\bigcup_{i\in U} \paras{t_i} & \text{if \(s= \{y_i \mid i\in U\} \subseteq Y_n\)
                  for some \(U \subseteq [n]\).}
%		\bigcup_{y_i \in t'} \paras{t_i} & \text{if \(t' \subseteq Y_{m'}\) is a sequence node}.
	\end{cases}
\end{align*}

%For all pairwise distinct symbols $\sigma_1\in\Delta^{(k_1)},\dots, 
%\sigma_n\in\Delta^{(k_n)}$ with $n\geq 1$ and
%$k_1,\dots,k_n\in\mathbb{N}$ and let $t_i$ for $i\in[n]$.
%Let $s\in T_\Delta$.
%Then $s[\![\sigma_i\leftarrow t_i\mid i\in[n]]\!]$ denotes the tree
%that is inductively defined as (abbreviating $[\![\sigma_i\leftarrow t_i\mid i\in[n]]\!]$ by
%$[\![\dots ]\!]$) follows:
%for $s=\sigma(s_1,\dots,s_k)$,
%if $\sigma\not\in\{\sigma_1,\dots,\sigma_n\}$ then $s[\![\dots]\!]=\sigma(s_1[\![\dots]\!],
%\dots,s_k[\![\dots]\!])$ and if $\sigma=\sigma_j$ for some $j\in[n]$ then
%$s[\![\dots]\!]=t_j[y_i\leftarrow s_i[\![\dots]\!]\mid i\in[k]]$.
%\item
Let $\gamma_1^{(k_1)},\dots,\gamma_n^{(k_n)}\in\Gamma$, $n\geq 1$ be pairwise different symbols
%Let $k_i=rank_\Gamma(\gamma_i)$ for $i\in[n]$
and assume now that 
$t_i\in T_\Gamma(Y_{k_i}\cup\mathcal{P}(Y_{k_i}))$ for $i\in[n]$
and that $s\in T_\Gamma(Y_n)$. 
The \emph{special second-order substitution} 
$[\![\gamma_i \leftarrow t_i\mid i\in[n]]\!]\sp{}$
(for short $[\![.]\!]\su$) applied to $s$ is
%inductively defined as 
%if $s=\gamma(s_1, \dots, s_k)$
%with $\gamma\not\in\{\gamma_1,\dots,\gamma_n\}$
%then $s[\![.]\!]\su = 
%\gamma(s_1[\![.]\!]\su, \dots, s_k[\![.]\!]\su)$, 
%if $s=y_j$ for $j\in [n]$ then $s[\![.]\!]\su = s$, 
%and if $s=\gamma_i$ for $i\in[n]$ then
%$s[\![.]\!]\su =  t_i
%[y_j \leftarrow t_j[\![.]\!]\su \mid j \in [n]]\su$. 
inductively defined as:
\begin{align*}
s[\![.]\!]\su &=
\begin{cases}
t_i
[y_j \leftarrow s_j[\![.]\!]\su \mid j \in [k_i]]\su
&
\text{if $s=\gamma_i(s_1, \dots, s_{k_i})$ for $i\in[n]$}
\\
\gamma(s_1[\![.]\!]\su, \dots, s_k[\![.]\!]\su)
& \text{if $s=\gamma(s_1, \dots, s_k)$
with $\gamma\not\in\{\gamma_1,\dots,\gamma_n\}$}
\\
s
&
\text{if $s=y_j$ for $j\in [n]$.}
\end{cases}
\end{align*}
%if $s=\gamma(s_1, \dots, s_k)$
%with $\gamma\not\in\{\gamma_1,\dots,\gamma_n\}$
%then $s[\![.]\!]\su = 
%\gamma(s_1[\![.]\!]\su, \dots, s_k[\![.]\!]\su)$, 
%if $s=y_j$ for $j\in [n]$ then $s[\![.]\!]\su = s$, 
%and if $s=\gamma_i$ for $i\in[n]$ then
%$s[\![.]\!]\su =  t_i
%[y_j \leftarrow t_j[\![.]\!]\su \mid j \in [n]]\su$. 

For all sets $Z \subseteq Y_m$ such that no $Y$-node in 
$t[\![.]\!]\su$ intersects $Z$, we define the $Z$-skeleton 
$\lcop{t[\![.]\!]\su}_Z$ of $t[\![.]\!]\su$ inductively as 
before, with a special case for $Y$-nodes: for all $Y$-nodes $S$ we 
have $\lcop{S}_Z = S \subseteq Y_m \setminus Z$. 
%\end{itemize}
\end{definition}

%Old version of the end of the definition of compatibility of rhs
%To define this notion of compatibility, we consider the second-order 
%substitution of a state call (in the right-hand-side of a rule) with the 
%state's skeleton, which induces a 
%first-order substitution of the parameters in the skeleton. We take such 
%first-order substitutions to apply \emph{within} sequence nodes of the skeleton 
%so that, given a sequence node $S=\{y_1,y_3\}$: 
%$S[y_i \leftarrow t_i]_{i\in [3]} = \{t_1,t_3\}$. 
%We assume given an mttr $M$ as before, with the condition that 
%$M$ is \emph{nondeleting}, which 
%means that all parameters of a state are used to build the state's output.
%
%\begin{definition}\label{def:meta-skeleta}
%For all states $q \in Q^{(m)}, q'\in Q^{(m')}$, $\sigma\in\Sigma^{(k)}$, 
%and $p_1,\dots,p_k\in P$, and noting $p=h(\sigma(p_1, \dots, p_k))$ and  
%$t=\text{rhs}_M(q,\sigma,\<p_1,\dots, p_k>)$, any state call $\<q',x_i>$ 
%in $t$ is \emph{compatible} with a set 
%$Z \subseteq Y_m$ if, for all $s_i \in L_{p_i}$, the sequence nodes in 
%$t[\![\<q',x_i>\leftarrow \lcop{M_{q'}(s_i)}_{Y(q',p_i)}]\!]$ contain no 
%parameters from $Z$. 
%\end{definition}

The special first-order substitution is
the same as the normal one
except that it gives special treatment to \(Y\)-nodes
which is replaced by \(Y\)-nodes contains all parameters occurring 
in trees to be substituted for the parameters in the original \(Y\)-nodes.
The special second-order substitution is
the same as the normal one
except that the special first-order substitution is applied
for each involved first-order substitution.

\begin{lemma}\label{lm:rhs}
Let $M$ be a nondeleting mttr as before.
Let $q\in Q$, $\sigma\in\Sigma^{(k)}$, and
$p_1,\dots,p_k\in P$. Let $p=h(\sigma(p_1, \dots, p_k))$ and 
$t=\text{rhs}_M(q,\sigma,\<p_1,\dots, p_k>)$. 
Let $s_1 \in L_{p_1}, \dots, s_k \in L_{p_k}$. By $[\![.]\!]^{\$}$ we
denote the substitution $[\![\<q',x_i>\leftarrow \lcop{M_{q'}(s_i)}_{Y(q',p_i)} \mid q'\in Q, 
i \in [k]]\!]^{\$}$ and by $[\![M]\!]$ we denote 
$[\![\<q',x_i>\leftarrow M_{q'}(s_i) \mid q'\in Q, i \in [k]]\!]$.
\begin{enumerate}
\item[(1)] If $y\in Y(q,p)$ and $y$ occurs in $t$ in the $j$-th argument 
of a node $\<q',x_i>$ for $q' \in Q$ and $i\in [k]$, then $y_j \in Y(q',p_i)$. 
%If $y\in Y(q,p)$ and $y$ occurs in  $t_j$ ($j\in[m]$), then $y_j\in Y(q',p_i)$.
\item[(2)] No $Y$-node in $t[\![.]\!]\su$ intersects $Y(q,p)$. 
%Any state call $\<q',x_i>$ in $t$ is compatible with $Y(q,p)$. 
\item[(3)] $\lcop{t[\![.]\!]\su}_{Y(q,p)} = \lcop{t[\![M]\!]}_{Y(q,p)}$
%For all input trees $s_i\in L_{p_i}$, we have: ~~~~
%$\lcop{t[\![\dots]\!]}_{Y(q,p)} = \lcop{t[\![\lcop{\dots}]\!]}_{Y(q,p)}$ \\
%where $[\![\dots]\!]$ denotes $[\![\<q',x_i>\leftarrow M_{q'}(s_i)]\!]$ \\
%and $[\![\lcop{\dots}]\!]$ denotes $[\![\<q',x_i>\leftarrow \lcop{M_{q'}(s_i)}_{Y(q',p_i)} ]\!]$. 
\end{enumerate}
\end{lemma}
\begin{proof}
If some $y_j\notin Y(q',p_i)$ then $\lcop{M_{q'}(L_{p_i})}_{y_j}$ is 
infinite and, if $y$ occurs in $t_j$ ($j\in[m]$), then 
$\lcop{M_{q}(L_{p})}_{y}$ is also infinite and $y\notin Y(q,p)$. 
So~(1)~holds. 

%(2)~is a consequence of~(1).
%Alternative proof of (2):
If $y \in Y(q,p)$ occurs in a $Y$-node of $t[\![.]\!]\su$, 
then it occurs in $t$ in the $j$-th argument of a node $\<q',x_i>$ with 
$y_j \notin Y(q',p_i)$, which contradicts~(1). 
So~(2)~holds. 

%As a consequence of~(1), parameter nodes and inner nodes are identical in 
%$\lcop{t[\![\dots]\!]}_{Y(q,p)}$ and 
%$\lcop{t[\![\lcop{\dots}]\!]}_{Y(q,p)}$. 
%Sequence nodes are also identical as a consequence of Lemma~\ref{lm:nd}(2). 
%Therefore~(3)~holds. 
%Alternatice proof of (3)
%%% Original proof FROM HERE %%%%%%%%%%%%%%%%%%%%%%%%%%%%%%%%%
%Both $\lcop{t[\![M]\!]}_{Y(q,p)}$ and 
%$\lcop{t[\![.]\!]\su}_{Y(q,p)}$ contain three types of nodes: 
%parameter nodes of the form $y\in Y(q,p)$, inner nodes and $Y$-nodes. 
%As a consequence of~(1), paths to parameters nodes are identical in 
%$\lcop{t[\![M]\!]}_{Y(q,p)}$ and $\lcop{t[\![.]\!]\su}_{Y(q,p)}$, 
%and the same is true of the inner nodes along such paths. 
%$Y$-nodes are also identical as a consequence of Lemma~\ref{lm:nd}(2) and 
%of our definition of special second-order substitutions. 
%So~(3)~holds. 
%%% Original proof TO HERE %%%%%%%%%%%%%%%%%%%%%%%%%%%%%%%%%

The statement~(3) is proved by induction on \(t\).
The cases of \(t=y_j\) and \(t=\gamma(t_1,\dots,t_n)\) are easy.
In the case of \(t=\<q',x_i>(t_1,\dots,t_m)\), we have
\begin{align*}
\lcop{t[\![.]\!]\su}_{Y(q,p)}
&=
\lcop{
\lcop{M_{q'}(s_i)}_{Y(q',p_i)}[y_j\leftarrow t_j[\![.]\!]\su\mid j\in[m]]\su
}_{Y(q,p)}
\\&=
\lcop{
\lcop{M_{q'}(s_i)}_{Y(q',p_i)}[y_j\leftarrow t_j[\![M]\!]\mid j\in[m]]\su
}_{Y(q,p)}
\\&=
\lcop{
M_{q'}(s_i)[y_j\leftarrow t_j[\![M]\!]\mid j\in[m]]
}_{Y(q,p)}
\\&=
\lcop{t[\![M]\!]}_{Y(q,p)}\text.
\end{align*}
%where the induction hypothesis and Lemma~\ref{lm:nd}(2) are used.
%
%Another alternatice proof of (3)
%To prove~(3)~we look at the $Y(q,p)$-skeleta of $t[\![\dots]\!]$ and 
%$t[\![\lcop{\dots}]\!]$. Those contain three types of nodes: 
%parameter nodes of the form $y\in Y(q,p)$, inner nodes (which are along 
%the paths to parameter nodes), and sequence nodes. Paths to parameters nodes 
%are identical in $t[\![\dots]\!]$ as in $t[\![\lcop{\dots}]\!]$ and so 
%are the inner nodes along such paths. Sequence nodes are also identical as a 
%consequence of Lemma~\ref{lm:nd}(2). 
%\qed
\end{proof}

\noindent
Statement of the Lemma~\ref{lm:corr}:
Let $M$ be a nondeleting mttr and $N=\pi(M)$ be the mttr of
Definition~\ref{df:pi}, both with the tuples as in that definition. 
Let $s\in T_\Sigma$ with $\hat{h'}(s)=(p,\varphi)$.
Then
\begin{enumerate}
\item[(1)] $p=\hat{h}(s)$,
\item[(2)] $\forall q\in F_p$: $\varphi(q)=\lcop{M_q(s)}_{Y(q,p)}$, 
\item[(3)] $\forall q\in Q$: $N_q(s)=M_q(s)$,
\item[(4)] $\forall q\in F_p$ and $u\in V(t)$ with $t=\varphi(q)$ and
$t/u=\{y_{j_1},\dots,y_{j_n}\}$ with \\
$j_1<\cdots <j_n$:
$N_{[q,p,t,u]}(s)=M_q(s)/u[y_{j_\nu}\leftarrow y_\nu\mid\nu\in[n]]$, and
\item[(5)] the mttr $N$ is nondeleting.
\end{enumerate}

\begin{proof}
All the statements are proven by induction on the structure of $s$.
Let $s=\sigma(s_1,\dots,s_k)$ with $\sigma\in\Sigma^{(k)}$,
$k\geq 0$, and $s_1,\dots,s_k\in T_\Sigma$.
For $i\in[k]$ let $\hat{h'}(s_i)=(p_i,\varphi_i)$.
By the definition of $h'$, $p=h_\sigma(p_1,\dots,p_k)$, which 
is equal to $\hat{h}(s)$.
Thus, Statement~(1) holds.
For Statement~(2) let $q\in F_p$: 
Then $\varphi(q)$ is defined as 
$\lcop{\zeta[\![\varphi_i]\!]\su}_{Y(q,p)}$ where 
$\zeta=\text{rhs}_M(q,\sigma,\<p_1,\dots,p_k>)$
and
$[\![\varphi_i]\!]\su$ denotes the special substitution
$[\![ \< q',x_i>\leftarrow \varphi_i(q')\mid q'\in F_{p_i}, i\in[k]]\!]\su$.
By induction, $\lcop{\zeta[\![\varphi_i]\!]\su}_{Y(q,p)}$ equals
$\lcop{\zeta[\![ \< q',x_i>\leftarrow \lcop{M_{q'}(s_i)}_{Y(q',p_i)}
\mid q'\in F_{p_i}, i\in[k]]\!]\su}_{Y(q,p)}$.
By Lemma~\ref{lm:rhs}(3) the latter equals
$\lcop{\zeta[\![ \< q',x_i>\leftarrow M_{q'}(s_i)
\mid q'\in F_{p_i}, i\in[k]]\!]}_{Y(q,p)}
=\lcop{M(s)}_{Y(q,p)}$.

We now prove Statement~(3).
Let $q\in Q$.
Then $N_q(s)=\zeta[\![ . ]\!][\![N]\!]$, 
where 
$\zeta=\text{rhs}_M(q,\sigma,\ab\<p_1,\dots,p_k>)$,
$[\![ . ]\!]$ is the substitution as in the construction, and
$[\![N]\!]=
[\![ \<r,x_i>\leftarrow N_r(s_i)\mid r\in Q',i\in[k] ]\!]$.
By the induction hypothesis of Statement~(2), we can replace
$\varphi_i(q')$ by $\lcop{M_{q'}(s_i)}_{Y(q',p_i)}$ in the
substitution $[\![ . ]\!]$. This gives
\begin{multline*}
\zeta
[\![ \<q',x_i>\leftarrow \lcop{M_{q'}(s_i)}_{Y(q',p_i)}
[
u'\leftarrow [q',p_i,\varphi_i(q'),u'](y_{j_1},\dots,y_{j_n})\mid \\
\varphi_i(q')/u'=\{y_{j_1},\dots,y_{j_n}\},
j_1<\cdots <j_n]
\mid q'\in F_{p_i},
i\in[k] ]\!]
[\![ N ]\!].
\end{multline*}
This can be written as 
$\zeta[\![.]\!][\![ H ]\!] [\![ Q ]\!]$,
where
$[\![ H ]\!]=[\![\< q',x_i>\leftarrow N_{q'}(s_i)\mid q'\in H,i\in[k] ]\!]$
and
$[\![ Q ]\!]  = 
[\![\< q',x_i>\leftarrow N_{q'}(s_i)\mid q'\in (Q \setminus F_{p_i}),i\in[k] ]\!]$.
By induction of Statement~(4) the substitution $[\![ H]\!]$ replaces the 
subtree $[q',p_i,\varphi_i(q'),u'](y_{j_1},\dots,y_{j_n})$ by the tree 
$M_{q'}(s_i)/u[y_{j_\nu}\ab\leftarrow y_\nu\mid\nu\in[n]]
[y_\nu\leftarrow y_{j_\nu}\mid\nu\in[n]]=M_{q'}(s_i)/u$. 
Thus we obtain:
\begin{multline*}
\zeta
[\![ \<q',x_i>\leftarrow \lcop{M_{q'}(s_i)}_{Y(q',p_i)}
[
u'\leftarrow M_{q'}(s_i)/u'\mid 
u'\in\mathcal{U}(\lcop{M_{q'}(s_i)}_{Y(q',p_i)})]\\
\mid q'\in F_{p_i},
i\in[k] ]\!] 
[\![ Q ]\!]
\end{multline*}
By Lemma~\ref{lm:nd} (for $Z=Y(q',p_i)$ and $t=M_{q'}(s_i)$)
the tree on the right of the arrow
in the leftmost second-order substitution equals $M_{q'}(s_i)$.
We have:
\[
\zeta
[\![ \<q',x_i>\leftarrow M_{q'}(s_i)\mid q'\in F_{p_i},i\in[k] ]\!] 
[\![\< q',x_i>\leftarrow N_{q'}(s_i)\mid q'\in Q \setminus F_{p_i},i\in[k] ]\!].
\]
By induction of Statement~(3), $N_{q'}(s_i)=M_{q'}(s_i)$ for
$q\in Q \setminus F_{p_i}$. This gives us exactly $M_q(s)$, by the definition
of the semantics of mttrs. 
Thus,
\begin{equation}\label{eq:MN}
N_q(s)=\zeta[\![ . ]\!][\![N]\!]=M_q(s).
\end{equation}
This concludes the proof of Statement~(3).

We now prove Statement~(4).
Let $q\in F_p$ and $u\in V(t)$ with $t=\varphi(q)$ 
and $t/u\subseteq Y$.
By the definition of the rules for the helper states, 
$N_{[q,p,t,u]}(s)=(\zeta[\![ . ]\!])/u[\![ N ]\!][y]$
where $t/u=\{y_{j_1},\dots,y_{j_n}\}$, $j_1<\cdots <j_n$, and 
$[y]=[y_{j_\nu}\leftarrow y_\nu\mid\nu\in[n]]$.
It follows from Lemma~\ref{lm:rhs}(1) that if $\<q',x_i>$ occurs in 
$\zeta=\text{rhs}_M(q,\sigma,\langle p_1,$ $\dots,p_k\rangle)$
and $q\in F_p$, then $q'\not\in Q \setminus F_{p_i}$.
Hence, every proper ancestor
$v$ of $u$ is labeled by a symbol in $\Delta$, i.e.,
$(\zeta[\![.]\!][y])[v]\in\Delta$. 
This implies that we can move the ``$/u$'' operation of 
taking the subtree at node $u$
to the right (after the application of the substitution $[\![ N ]\!]$)
in the above displayed formula.
We obtain 
$\zeta[\![ . ]\!][\![ N ]\!]/u[y]$.
By the right equation in Formula~\ref{eq:MN}, this equals
$M_q(s)/u[y]$.

To prove Statement~(5), let $q\in Q^{(m)}$, $m\geq 0$.
Then 
\[
\zeta'=\text{rhs}_N(q,\sigma,\< (p_1,\varphi_1),\dots,(p_k,\varphi_k)>)=
\zeta[\![.]\!],
\]
where 
$\zeta=\text{rhs}_M(q,\sigma,\<p_1,\dots,p_k>)$ and
$[\![.]\!]$ is as before. 
By Statement~(2), $[\![.]\!]$ substitutes occurrences of 
$\<q',x_i>$ with $i\in[k]$ and $q'\in F_{p_i}$ by 
the tree $\lcop{M_{q'}(s_i)}_{Y(q',p_i)}$ in which leaves
labeled by $Z\subseteq Y_m$ are replaced by $\<q_H,x_i>(y_{j_1},\dots,
y_{j_n})$ with $Z=\{y_{j_1},\dots,y_{j_n}\}$.
By Lemma~\ref{lm:nd}(2) this implies that 
$y_j$ occurs in $\zeta'$ for each $j\in[m]$.
%\qed
\end{proof}

\subsection*{Proof of Lemma~\ref{lm:dec}}

Statement of the lemma:
Let $M=(Q,P,\Sigma,\Delta,q_0,R,h)$ be an mttr and let
$q\in Q^{(m)}$, $m\geq 1$, $j\in[m]$, and $p\in P$.
It is decidable whether or not 
$\lcop{M_q(L_p)}_{\{y_j\}}$ is finite.
In case of finiteness, 
$\lcop{M_q(L_p)}_{\{y_j\}}$ can be constructed.

\begin{proof}
Let $Z=\{y_j\}$.
We now consider symbols in $Y_m$ as rank zero symbols.
It is straightforward to construct a  top-down tree transducer with look-ahead $M_Z$ which outputs
$\lcop{t}_Z$ for input trees $t\in T_{\Delta\cup Y_m}$; note that
$T_{\Delta\cup Y_m}=T_\Delta(Y_m)$ because symbols in $Y_m$ are now considered
as symbols of rank zero.
The transducer $M_Z$ computes in its look-ahead $h'$ the set of parameters of
the input tree, i.e.\ $h'(t)=\paras{t}$ for every $t\in T_{\Delta\cup Y_m}$. 
The transducer $M_Z$ (which consists of a single state only) outputs $\paras{t}$
as soon as $\paras{t}\cap Z=\emptyset$.
%The details can be found in the Appendix.

Formally, $M_Z = (\{q_1^{(0)}\},P',\Delta\cup Y_m,\Delta',q_1,R',h')$ where
$P'={\cal P}(Y_m)$ and
$\Delta'=\Delta\cup \{ S^{(0)}\mid S\in P'\}\cup Y_m$.
For $y\in Y_m$ let $h'_y()=\{y\}$ and for
$a\in\Delta^{(0)}$ let $h'_a()=\emptyset$.
Further, for $\delta\in\Delta^{(k)}$, $k\geq 1$, and 
$S_1,\dots,S_k\in P'$ let 
$h'_\delta(S_1,\dots,S_k)=\bigcup_{i\in[k]}S_i$.
For $a\in\Delta^{(0)}$ let the rule
$\< q_1,a>\to\emptyset$ be in $R$.
Let $y\in Y_m$.
If $y\in Z$ then let the rule
$\< q_1,y>\to y$ be in $R$ and otherwise let the rule
$\< q_1,y>\to \{y\}$ be in $R$.
For $\delta\in\Delta^{(k)}$ with $k\geq 1$ 
and $S_1,\dots,S_k\in P'$
we define the rule
$\< q_1,\delta(x_1:S_1,\dots,x_k:S_k)>\to\zeta$ where $\zeta$ is defined as:
\[
\begin{array}{lcll}
  \zeta & = &
\left\{ 
  \begin{array}{ll}
  S & \text{if }S= (\bigcup_{i\in[k]} S_i)\cap Z=\emptyset\\
  \delta(\< q_1,x_1>,\dots,\<q_1,x_k>) & \text{otherwise.}
  \end{array}
\right. 
\end{array}
\]

\noindent
\textbf{Claim.}\quad
For every $t\in T_{\Delta\cup Y_m}$, $h'(t)=\paras{t}$ and
$M_Z(t)=\lcop{t}_Z$.

\medskip

It is straightforward to prove this claim by induction on the structure of $t$.
By the Claim,
$M_{\{y_j\}}(M_q(L_p)) = \lcop{M_q(L_p)}_{\{y_j\}}$.
The tree language $L_p$ can be represented by a partial nondeterministic
top-down tree transducer that realizes the identity on trees in $L_p$
(and is undefined otherwise; its rules are obtained by reading the
definitions of $h$ from right to left). 
In this way $M_{\{y_j\}}(M_q(L_p))$ is represented
so that its finiteness is decidable by Proposition~\ref{prop:finite}
(and in case of finiteness the set can be constructed).
%\qed
\end{proof}

\subsection*{Proof of Lemma~\ref{lm:decidable}}

Statement of the lemma:
Let $M$ be an mttr.
Then 
(1)~it is decidable whether or not $M$ is finite nesting and
(2)~it is decidable whether or not $M$ is finite yield nesting.

\smallskip

\begin{proof}
Let $M=(Q,P,\Sigma,\Delta,q_0,R,h)$.
We use the extension 
$\widehat{M}=(\hat{Q},P,\hat{\Sigma},\hat{\Delta},q_0,$ $\hat{R},h)$
of $M$ with input trees in $s\in T_\Sigma(P)$ which contain
(1)~exactly one or
(2)~arbitrarily many occurrences of elements of $P$.
We then use a nondeterministic top-down tree transducer $N$ which chooses
any path in the tree $\widehat{M}(s)$ and outputs only the elements from
$\< Q,P>$ on that path, now seen as unary symbols.
The resulting output language $N(\widehat{M}(T_\Sigma))$ is finite if and only
if $M$ is (1)~fnest or (2)~fynest.

Formally, 
$N=(\{q_1^{(0)}\},\hat{\Delta},\Gamma,q_1,R')$ where
$\Gamma=\< Q,P>\cup\{e^{(0)}\}$.
For every $\delta\in\Delta^{(k)}$, $k\geq 1$, and $i\in k$
we let the rule
$\<q_1,\delta(x_1,\dots,x_k)>\to\<q_1,x_i>$ be in $R'$.
For every $\delta\in\Delta^{(0)}$ we let the rule
$\<q_1,\delta>\to e$ be in $R'$.
For every $\<q,p>\in \<Q,P>^{(m)}$, $m\geq 1$, and $i\in[m]$
we let the rule
$\<q_1,\langle q,p\rangle(x_1,\dots,x_m)>\to \<q,p>(\<q_1,x_i>)$ be in $R'$.
For every $\<q,p>\in\<Q,P>^{(0)}$ we let the rule
$\<q_1,\langle q,p\rangle>\to \<q,p>$ be in $R'$.
It is straightforward to show (by induction on the structure of $s$), that
$N(\widehat{M}(T_\Sigma(P)))$ is finite if and only if $M$ is fynest.
Let $L$ be the set of trees in $T_\Sigma(P)$ which contain exactly one occurrence
of an element of $P$. 
It is straightforward to show (by induction on the structure of $s$), that
$N(\widehat{M}(L))$ is finite if and only if $M$ is fnest.
%\qed
\end{proof}

\subsection*{Proof of Lemma~\ref{lm:easy}}

Statement of the lemma:
Let $M$ be an mttr.
(1)~If $M$ is finite nesting, then it is of linear size-to-height increase.
(2)~If $M$ is finite yield nesting, then it is of linear height increase. 

\begin{proof}
Informally, we can understand this lemma by looking at a given path $O$ in an output tree and, using origin semantics, at how many nodes along this path have their origin in different parts of the input tree. 

For~(1), the finite nesting property gives a bound $c$ on the number of state calls to a single input node, nested along path $O$. Intuitively, noting $\text{mhr}$ the maximum height of the right-hand side of a rule, $c.\text{mhr}$ is a bound on the number of output nodes along path $O$ with their origin in a single input node. This bound clearly implies that the height of the output (maximum number of nodes on a path) is linearly bound by the size of the input. 

For~(2), instead of looking at a single input node, we look at all the input nodes at a given depth $d$ in the input. The finite yield nesting property implies a bound $c$ on the nesting (along a path $O$) of state calls to input nodes of depth $d$. Each such call may produce at most $\text{mhr}$ nodes along path $O$ with their origin in a node of depth $d$. So $c.\text{mhr}$ is a bound for the number of nodes along path $O$ with their origin in a node of depth $d$. 
%Note that these bounds only work because we look at state calls along an output path, because states of MTTs can copy other state calls when they appear in their parameters, but cannot copy them vertically (i.e.\ two such copied state calls cannot appear nested along a same output path). 

Formally, we apply $\widehat{M}$ to a tree $t\in T_\Sigma(P)$. We modify $t$ by substituting nodes in $P$, and we bound the growth of the height of $\widehat{M}(t)$ for each substitution. We will conclude by stating that any input tree $s \in T_\Sigma$ can be built by successive substitutions, and so the height of the output is linearly bound by the number of substitutions (which will be the size of $s$ for~(1), and the height of $s$ for~(2)). Let $M=(Q,P,\Sigma,\Delta,q_0,R,h)$ and let $\text{mhr}$ be the maximum height of the right-hand side of any rule in $R$. Let $s$ be a fixed tree in $T_\Sigma$.

To prove~(1), consider $U$ an arbitrary set of pairwise independent (i.e.\ not being descendants of each other) nodes of a fixed input tree $s \in T_\Sigma$. Let $s'=s[u\leftarrow h(s/u)\mid u\in U]$, let $u \in U$, 
%$u\in V(s')$ with $s'/u\in P$, 
and $\sigma=s[u]\in\Sigma^{(k)}$ with $k\geq 0$. Let $c$ be a nesting bound for $M$, then, along any output path in $\widehat{M}(s')$, there are at most $c$ state calls $\<q,s'/u>$ with origin $u$ in $s'$. Then $\widehat{M}(s'[u\leftarrow\sigma(h(s/u1),\dots,h(s/uk))])$ is obtained by replacing such state calls with the corresponding right-hand side of rules, which implies that: $\he{\widehat{M}(s'[u\leftarrow\sigma(h(s/u1),\dots,h(s/uk))])} \leq c\cdot\text{mhr} + \he{\widehat{M}(s')}$. 
The tree $s \in T_\Sigma$ can be obtained from the tree $h(s) \in T_\Sigma(P)$ by $|s|$ such substitutions. 
The height of $\widehat{M}(h(s)) = \<q_0,h(s)>$ is $0$. So $\he{M(s)} \leq c\cdot\text{mhr} \cdot |s|$, so $M$ of linear size-to-height increase. 

To prove~(2), for $i\in[\he{s}]$, let $U_i$ be the set of nodes at depth $i$ in $s$, and consider $s_i$ be the tree obtained from $s$ by replacing all nodes $u \in U_i$ by $h(s/u)$. Let $c$ be a yield nesting bound for $M$, then, along any output path $O$ in $\widehat{M}(s_i)$, there are at most $c$ state calls $\<q,s_i/u>$ with $u \in U_i$. 
%So we can substitute all the state calls on input subtrees of depth $i$ at the same time, and increase the height of the output by less than $c\cdot \text{mhr}$.
Then $\widehat{M}(s_{i+1}) = \widehat{M}(s_i[u\leftarrow \sigma_u(h(s/u1),\dots,h(s/uk)) \mid u \in U_i])$ is obtained by replacing these state calls in $\widehat{M}(s_i)$ with the corresponding right-hand side of rules, so $\he{\widehat{M}(s_{i+1})} \leq c\cdot\text{mhr} + \he{\widehat{M}(s_i)}$.
%Therefore: $\he{\widehat{M}(s_{i+1})} \leq c\cdot\text{mhr} + \he{\widehat{M}(s_i)}$. 
By applying this $\he{s}+1$ times, starting from $s_0$, we obtain that the height of $M(s)$ is $\leq c\cdot\text{mhr} \cdot(\he{s}+1)$. So $M$ is of linear-height-increase. 

\end{proof} 
%\vspace{-1mm}

\end{document}